\documentclass[twoside]{amsart}
\usepackage{amsmath}
\usepackage{amscd}
\usepackage{amsfonts}
\usepackage{setspace}
\usepackage{ifthen}

\newcommand{\commentout}[1]{}

{\begin{enumerate}{}%
{

}
}%
{\end{enumerate}}

\newcommand{\tighten}{\setlength{\arraycolsep}{1pt}}

\def\today{\ifcase\month\or
 January\or February\or March\or April\or May\or June\or
 July\or August\or September\or October\or November\or December\fi
 \space\number\day, \number\year}

\newcommand{\CC}{\ensuremath{\mathbb{C}}}
\newcommand{\FF}{\ensuremath{\mathbb{F}}}

\newcommand{\RR}{\ensuremath{\mathbb{R}}}
\newcommand{\ZZ}{\ensuremath{\mathbb{Z}}}
\newcommand{\PP}{\ensuremath{\mathbb{P}}}

\newcommand{\mytexorpdfstring}[2]{\texorpdfstring{#1}{#2}}

\newtheorem{theorem}{Theorem}
\newtheorem{definition}{Definition}
\newtheorem{lemma}{Lemma}
\newtheorem{obs}{Observation}
\newtheorem{prop}{Proposition}
\newtheorem{claim}{Claim}

\renewcommand{\eqref}[1]{(\ref{eq:#1})}
\newcommand{\thmref}[1]{Theorem~\ref{thm:#1}}
\newcommand{\propref}[1]{Proposition~\ref{prop:#1}}
\newcommand{\secref}[1]{Section~\ref{sec:#1}}

\newcommand{\lemmaref}[1]{Lemma~\ref{lemma:#1}}
\def\marginnote#1{\marginpar{\ifodd\value{page}
\rlap{#1}
\else\hskip\marginparwidth\llap{#1}
\fi
}}
\newcommand{\gc}[1]{{\bf [[Note: {#1}]]}}
\newcommand{\loosen}{\setlength{\arraycolsep}{3pt}}
\newcommand{\matrixijkl}[4]{\ensuremath{%
 \left(\loosen{\begin{array}{@{}rc@{}} #1 & #2 \\ #3 & #4 \end{array}}\right)}}

\newcommand{\action}{{action}{}}
\newcommand{\actions}{actions{}}
\renewcommand{\epsilon}{\varepsilon}
\newcommand{\enm}[1]{\ensuremath{#1}}
\newcommand{\superijk}[3]{\enm{{#1}^{#2|#3}}}
\newcommand{\mij}[2]{\superijk{M}{#1}{#2}}
\newcommand{\rrij}[2]{\superijk{E}{#1}{#2}}
\newcommand{\rrpq}{\superijk{\RR}{p}{q}}

\renewcommand{\aa}{\enm{\mathcal{A}}}
\newcommand{\sa}{\enm{S\aa}}
\newcommand{\bb}{\enm{\mathcal{B}}}
\renewcommand{\gg}{\enm{\mathcal{G}}}
\newcommand{\oi}[1]{\enm{\mathcal{O}^{(#1)}}}
\newcommand{\mm}{\enm{\mathcal{M}}}
\newcommand{\g}{\enm{\mathfrak{g}}}

\newcommand{\cdibar}[1]{\enm{\bar{\mathcal{D}}_{\dot{#1}}}}

\newcommand{\diraci}[1]{\enm{{#1}\hskip-.65em /}}

\newcommand{\odiraca}{\diraci{D}}
\newcommand{\im}{\mathrm{Im}}

\newcommand{\ip}[2]{\enm{\left<#1,#2\right>}}

\renewcommand{\bar}[1]{\overline{#1}}

\renewcommand{\mij}[2]{\enm{M^{{#1}|{#2}}}}
\newcommand{\plphi}{\enm{\hat{\phi}}}
\newcommand{\plpsi}{\enm{\hat{\psi}}}
\newcommand{\plf}{\enm{\hat{F_{A}}}}
\newcommand{\f}{\enm{\mathcal{F}}} 

\newcommand{\spin}{\enm{\mathrm{\it Spin}}}

\newcommand{\di}[1]{\enm{\partial_{#1}}}
\newcommand{\dicj}[1]{\enm{\bar{\partial}_{\dot{#1}}}}
\newcommand{\dieven}[2]{\enm{\partial_{#1\dot{#2}}}}
\newcommand{\ditil}[1]{\enm{\tilde{\partial}_{#1}}}

\newcommand{\cjsup}[2]{\enm{\bar#1{} ^{{\dot{#2 }}}}} 
\newcommand{\cjsub}[2]{\enm{\bar#1{} _{{\dot{#2 }}}}} 
\newcommand{\odi}[1]{\enm{D_{#1}}}
\newcommand{\odicj}[1]{\enm{\cjsub{D}{#1}}}
\newcommand{\oda}{\odi{a}}
\newcommand{\odacj}{\odicj{a}}
\newcommand{\oqi}[1]{\enm{Q_{#1}}}
\newcommand{\oqicj}[1]{\enm{\cjsub{Q}{#1}}}
\newcommand{\oqa}{\oqi{a}}
\newcommand{\oqacj}{\oqicj{a}}
\newcommand{\even}[3]{\enm{{#1}^{#2 \dot{#3}}}}
\newcommand{\onetwoij}[2]{\enm{#1^{(#2)}}}
\newcommand{\onetwoi}[1]{\onetwoij{#1}{i}}
\newcommand{\one}[1]{\onetwoij{#1}{1}}
\newcommand{\two}[1]{\onetwoij{#1}{2}}
\newcommand{\sut}{\enm{SU(2)}}
\newcommand{\sot}{\enm{SO(3)}}
\newcommand{\sof}{\enm{SO(4)}}
\newcommand{\uo}{\enm{U(1)}}
\newcommand{\pp}{\enm{\mathcal{P}}}
\newcommand{\dicov}[1]{\enm{\nabla_{#1}}}
\newcommand{\dievencov}[2]{\enm{\nabla_{#1\dot{#2}}}}

\newcommand{\odicov}[1]{\enm{\mathcal{D}_{#1}}}

\newcommand{\odicjcov}[1]{\enm{\bar{\mathcal{D}}_{\dot{#1}}}}

\newcommand{\ad}{\enm{\mathrm{ad}}}

\newcommand{\odijcov}[2]{\onetwoij{\odicov{#2}}{#1}}
\newcommand{\odijcjcov}[2]{\onetwoij{\odicjcov{#2}}{#1}}

\newcommand{\formsij}[2]{\ensuremath{\Omega^{#1}({#2})}}

\newcommand{\spl}{\enm{S^{+}}}
\newcommand{\sm}{\enm{S^{-}}}
\newcommand{\spm}{\enm{S^{\pm}}}
\newcommand{\spd}{\enm{(\spl)^{*}}}
\newcommand{\smd}{\enm{(\sm)^{*}}}
\newcommand{\eps}{\enm{\epsilon}}

\newcommand{\eij}[2]{\enm{e^{#1}_{#2}}}
\newcommand{\ei}[1]{\enm{e^{#1}_{+}}}
\newcommand{\eib}[1]{\enm{e^{#1}_{-}}}
\newcommand{\eil}[1]{\enm{e_{#1}^{+}}}
\newcommand{\eilb}[1]{\enm{e_{#1}^{-}}}
\newcommand{\eipm}[1]{\eij{#1}{\pm}}
\newcommand{\eilpm}[1]{\eij{\pm}{#1}}

\renewcommand{\epsilon}{\varepsilon}
\newcommand{\wdd}[2]{\enm{\partial_{x^{#1}}\wedge \partial_{x^{#2}}}}
\newcommand{\wdu}[2]{\enm{\partial_{x^{#1}}\otimes dx^{#2}}}

\newcommand{\opp}[1]{\enm{{#1}{}^{\mathrm{opp}}}}

\newcommand{\bothupperabc}[3]{\enm{{#1}^{#2 #3}}}
\newcommand{\bothlowerabc}[3]{\enm{{#1}_{#2 #3}}}
\newcommand{\upperlowerabc}[3]{\enm{{#1}^{#2}_{#3}}}
\newcommand{\dotbothupperabc}[3]{\bothupperabc{#1}{#2}{\dot{#3}}}
\newcommand{\dotbothlowerabc}[3]{\bothlowerabc{#1}{#2}{\dot{#3}}}
\newcommand{\upperdotlowerabc}[3]{\upperlowerabc{#1}{#2}{\dot{#3}}}

\renewcommand{\ell}[2]{\bothlowerabc{e}{#1}{#2}}
\newcommand{\euu}[2]{\bothupperabc{e}{#1}{#2}}
\newcommand{\eul}[2]{\upperlowerabc{e}{#1}{#2}}
\newcommand{\eudl}[2]{\upperdotlowerabc{e}{#1}{#2}}
\newcommand{\eldl}[2]{\dotbothlowerabc{e}{#1}{#2}}
\newcommand{\tudu}[2]{\dotbothupperabc{\theta}{#1}{#2}}
\newcommand{\pldl}[2]{\dotbothlowerabc{\partial}{#1}{#2}}

\newcommand{\dul}[2]{\upperlowerabc{D}{#1}{#2}}
\newcommand{\dudl}[2]{\upperdotlowerabc{\bar{D}}{#1}{#2}}
\newcommand{\dudlc}[2]{\upperdotlowerabc{\bar{\mathcal{D}}}{#1}{#2}}

\newcommand{\qul}[2]{\upperlowerabc{Q}{#1}{#2}}

\newcommand{\susols}{\cite{susols}}
\newcommand{\nab}{\enm{\nabla}}
\newcommand{\doc}{\enm{\mathcal{D}_{1}}}
\newcommand{\dzc}{\enm{\mathcal{D}_{0}}}
\newcommand{\qoc}{\enm{\mathcal{Q}_{1}}}
\newcommand{\qzc}{\enm{\mathcal{Q}_{0}}}
\newcommand{\dtc}{\enm{\mathcal{D}_{2}}}
\newcommand{\dd}{\enm{\mathcal{D}}}

\newcommand{\ff}{\enm{\mathcal{F}}}
\newcommand{\adp}{\enm{\ad\,P}}
\newcommand{\lieg}{\enm{\mathrm{Lie}\,\gg}}

\newcommand{\thom}{\enm{\mathcal{T}}}

\newcommand{\bianchioddoddeven}[6]{
    \begin{array}{ccccccc}
           & {#4}(\ff({#2}, {#3}))& - & 
             \ff([{#1},{#2}], {#3}) & + & 
             \ff({#2}, [{#1},{#3}]) & \\
        - & {#5}(\ff({#3}, {#1}))& + & 
             \ff([{#2},{#3}], {#1}) & + & 
             \ff({#3}, [{#2},{#1}]) & \\
        + & {#6}(\ff({#1}, {#2}))& - & 
             \ff([{#3},{#1}], {#2}) & - & 
             \ff({#1}, [{#3},{#2}]) & =0\\
    \end{array}
}
\newcommand{\bianchioddoddodd}[6]{
    \begin{array}{ccccccc}
           & {#4}(\ff({#2}, {#3}))& - & 
             \ff([{#1},{#2}], {#3}) & + & 
             \ff({#2}, [{#1},{#3}]) & \\
        + & {#5}(\ff({#3}, {#1}))& - & 
             \ff([{#2},{#3}], {#1}) & + & 
             \ff({#3}, [{#2},{#1}]) & \\
        + & {#6}(\ff({#1}, {#2}))& - & 
             \ff([{#3},{#1}], {#2}) & + & 
             \ff({#1}, [{#3},{#2}]) & =0\\
    \end{array}
}

\newcommand{\cbi}[1]{\enm{\bar{\chi}_{\dot{#1}}}}
\newcommand{\lbi}[1]{\enm{\bar{\lambda}_{\dot{#1}}}}

\title{A Supersymmetric Quantum Field Theory Formulation of the Donaldson
Polynomial Invariants}
\author{Gregory Langmead}
\date{October 21, 2002}
\begin{document}
\begin{abstract}
We construct a mathematical framework for twisted \( N=2 \)
supersymmetric topological quantum field theory on a 4-manifold. 
Supersymmetry in flat space is defined and the twist homomorphism is
constructed, giving us a supermanifold that is the total space of an
odd vector bundle over the even 4-manifold.  A special category of
connections on this space is defined and a decomposition into
so-called component fields is proved.  The twisted supersymmetric
action is computed, and the structure of the action, the
decomposition, and the action of a special odd vector field are all
shown to have a rich geometrical structure that was partially
interpred by Atiyah and Jeffrey. \cite{atiyahjeffrey}  
In short, the
action is an infinite-dimensional analogue of the Euler class of the
vector bundle of self-dual 2-forms over the space of connections mod
gauge.  This geometrical insight serves two purposes: first, it
motivates the study of anti-self-dual connections, intersection
theory, and the action of the group of gauge transformations, all of
which appear by themselves after the twist.  Secondly, it sets the
stage for an eventual proof of Witten's Conjecture, relating the
Donaldson and Seiberg-Witten invariants.  What we build here amounts
to a mathematical treatment of a physical treatment \cite{wittentqft}
of a mathematical construction of Donaldson. \cite{donaldsonpoly},
\cite{donaldsonkronheimer}.
\end{abstract}

\maketitle

\pagestyle{myheadings}
\markboth{GREGORY LANGMEAD}{A SUSY FORMULATION OF DONALDSON THEORY}
\section*{Introduction} The primary goal of this paper is to present an
alternative formulation of Donaldson theory \cite{donaldsonpoly},
\cite{donaldsonkronheimer}.  This will involve an exploration of
supersymmetry and an important variation thereof.  We will construct a
very special eight-dimensional vector bundle \( SX \) over a compact,
closed, simply connected riemannian four-manifold \( X \) that is the
direct descendent of \( N=2 \) supersymmetry in Euclidean space.  We
will examine the space of connections on a principal bundle over this
space, building on results for connections over super Euclidean space. 
This space of connections comes equipped with a vector field,
inherited from the supersymmetry algebra.  The structure of the space
together with the vector field is very rich and generalizes a
beautiful finite dimensional geometrical picture.  This geometry is
further reflected in the \action, a functional on superconnections
that is the analog for \( SX \) of a similar construct in super
Euclidean space.  What we gain from this framework is a set of
algebraic tools that are geared to one purpose: doing intersection
theory on the moduli space of anti-self-dual (ASD) connections modulo
the group of gauge transformations.  Without having introduced ASD
connections, or requiring that we divide by the gauge group, we see
that these objects and operations are natural in this supersymmetric
context.

The secondary goal of this paper is discussed in the final section. 
Once we have seen that the Donaldson invariants fit into a
supersymmetric quantum field theory framework, we can begin to address
Witten's Conjecture \cite{witten94}.  This is the famous unproven
relationship between the Donaldson and Seiberg-Witten 
(\cite{seibergwitten1}, \cite{morgan}) invariants. 
Witten's ``proof'' of this result used the celebrated breakthroughs he
obtained with Seiberg \cite{seibergwitten1} on \( N=2 \)
supersymmetric gauge theories in Minkowski space, just like the gauge theory we
consider here in Euclidean space.  We will see a brief sketch of their
proof, and attempt to point the way that leads from this paper to a
mathematical proof of their result.  We hope to convince the reader
that the enlargement of Donaldson's picture presented here is the
right place to begin to understand why Witten's conjecture is true,
and maybe why and how it was discovered in the first place.

The following outline of this document should aid the reader.  Super
Euclidean and super Minkowski space are made from spin bundles in
four dimensions, so we present these objects and the necessary volume
forms and metrics.  Super Euclidean space is the first important
object we will encounter.  This superspace has a special framing that
reflects the spin structure, and which we will use to construct
component fields of superfields.  This framing is the second focal
point of this paper.  Next we define a very specific category of
connections over superspace, called \emph{semi-constrained}.  The
semi-constrained condition comes from physics, where objects are
defined in terms of a dimensional reduction from six dimensions to
four.  In short, fully constrained connections are required to be flat
in the odd directions, whereas semi-constrained ones can have two
independent nonvanishing components of curvature in odd directions. 
Semi-constrained connections form the third focal center for this
work.  To study them, we describe the approach taken in physics.  That
is, we construct and prove an isomorphism between superconnections and
a different space, a product space of objects defined on the
underlying even principal bundle and even four-manifold.  We rely
heavily on the framing we constructed to define these ``component
fields.''  Finally, we put this all together and write down the \( N=2
\) superspace version of the Yang-Mills action, which involves all of
the component fields.  

Section 2 repeats much of this discussion for the \emph{twisted} 
picture.  The twist is a representation theoretic operation on the 
two copies of spinors we have in \( N=2 \) supersymmetry, that turns 
spinors into constants, 1-forms, and self-dual 2-forms.  It is here 
that self-duality enters for the first time, and it is directly from 
this isomorphism of representations that we are eventually led to 
consider the ASD equations.  We construct analogues 
of material from Section~1: the eight-dimensional bundle is now 
well-defined on any riemannian 4-manifold; the eight odd vector 
fields that formed our framing become three odd vector fields with 
bundle values; superconnections retain the same definition, though the 
nonvanishing odd-odd curvatures are described differently; the 
component fields have a more elegant and intrinsic definition, though 
we are careful to recognize that each new component field can be 
rewritten in a coordinate patch as its flat space counterpart.  Then 
we write down the action of one of the three odd vector fields on the 
space of superconnections.  This action forms part of an infinite-dimensional 
analogue of a beautiful finite-dimensional construction, which we take 
up in Section~3.

Section 3 is an introduction to two finite-dimensional geometrical 
constructions.  The first is a special formula for the Thom class of 
a vector bundle, first constructed by Mathai and Quillen \cite{mathaiquillen}.  The second 
is a form on the total space of a principal bundle that allows 
integration of forms on the base to take place on the total space 
instead. Both constructions have an algebraic flavor that helped 
physicists connect with physics.  We will see 
that the algebraic structure of both of these constructions 
is present on our 4-manifold in the form of: the space of 
superconnections, the action of the odd vector field, and the form of 
the twisted Yang-Mills action.  This would be enough to prove the 
field theoretic formulas for the Donaldson polynomial invariants that 
we write down, but for the fact that there are no theorems along the geometrical lines that work in finite dimensions.  However, 
we will prove the result another way, using formulas for linear and 
Gaussian path integrals that are formal but consistent with physical manipulations.

Finally, in the last section we introduce the reader to the issues
that led to this work, namely Witten's Conjecture relating the
Donaldson polynomial invariants to the Seiberg-Witten invariants.  We
will see that in the picture painted by modern physics, the quest for
an easier formulation of the polynomial invariants is \emph{completely
natural}.  The issue is that solving this problem is very hard. 
There is a physics proof, and any proper mathematical proof should 
address it, or at least parallel it.  And so we are led to ask for a 
mathematical formulation of the physical formulation of the 
invariants, a need this paper is designed to address.

\section{Introduction to supersymmetry}

\subsection{A few super preliminaries}

\label{sec:superprelims}
A super vector space is a \(\ZZ/2\ZZ\)-graded vector space
\[ V = V_0\oplus V_1.\]
The \emph{parity} of an element \( v\in V \), denoted \( \pi(v) \), 
is \( 0 \) if \( v\in V_0 \), in which case \( v \) is called 
\emph{even}, and is 1 if \( v\in V_1 \), in which case \( v \) is 
called \emph{odd}.
A morphism from \( V \) to \( W \) in this category is a 
grading-preserving linear transformation.  The \emph{parity reversal} 
of \( V \), denoted \( \Pi V \) is an isomorphism defined by
\begin{gather*}
   \left(\Pi V\right)_0 = V_1 \\
   \left(\Pi V\right)_1 = V_0.
\end{gather*}
Tensor products are defined using the tensor product of the 
underlying vector spaces, with grading given by
\[ (V\otimes W)_k = \oplus_{i+j=k}V_i\otimes W_j. \]
The departure from simply defining a category of graded spaces comes 
with the definition of the commutativity isomorphism
\[ V\otimes W \to W\otimes V \]
which we define to send
\begin{equation}
   v\otimes w \to (-1)^{\pi(v)\pi(w)}w\otimes v.
\end{equation}

If \( t_1,\ldots,t_p \) is a basis for \( V_0 \) and \( \theta_1,\ldots,\theta_q \) is a basis for \( V_1 \), the commutative \RR-algebra \( \RR[t_1,\ldots,t_p,\theta_1, \ldots, 
\theta_q] \) is defined to be
\[ S^*(t_1,\ldots,t_p)\otimes\wedge^*(\theta_1,\ldots\theta_q).\]
This should be thought of as a super version of the symmetric algebra
on a vector space, where skew commutativity of the \( \theta_i
\) is part of the underlying properties of the odd generators.  
The
space \rrpq\ is defined as the topological space \( \RR^p \) endowed
with a sheaf \( C^{\infty}(\RR^p)(\theta^1,\ldots\theta^q) \) of
commutative super \RR-algebras, freely generated over the sheaf \(
C^{\infty}(\RR^p) \) by the odd quantities \( \theta^1,\ldots\theta^q
\).  A \emph{super manifold} \( M \) is a topological space with a
sheaf of super \RR-algebras, that is locally isomorphic to \rrpq.  The
ideal generated by all odd functions on a supermanifold \( M \)
defines an even submanifold we will denote by \( M_{\mathrm{even}}, \)
where we use the usual algebro-geometric correspondence between ideals
and varieties given by the set of common zeros of the ideal.

A morphism from a supermanifold \( S \) to \rrpq\ can be identified 
with a set of \( p \) even functions and \( q \) odd functions on \( 
S \).  This definition can be worked up into a definition of maps 
between supermanifolds, and in particular to vector bundles and 
principal bundles.  If \( \pp\to M \) is a principal \sut\ bundle 
over a supermanifold \( M \), with fiber given by the even space \sut, 
then the restriction of this bundle to the even part \( 
M_{\mathrm{even}} \) of \( M \) is a usual principal bundle we will 
consistently denote by \( P \).

If \( E\to M \) is a vector bundle over a supermanifold, with fiber 
isomorphic to \rrpq, then there is another vector bundle we can form 
denoted \( \Pi E\to M \), which is parity reversed on each fiber.  In 
this case, the underlying even vector bundle has fiber isomorphic to 
\( \RR^q \), whereas the even vector bundle underlying \( E \) has 
fiber isomorphic to \( \RR^p \). 

There is a concept of integration over odd variables called 
\emph{Berezinian integration}.  To compute
\[ \int d\theta_{1}\cdots d\theta_{n} f(x_{1}, \ldots, x_{k}, 
\theta_{1}, \ldots, \theta_{l}) \]
we expand \( f \) into a power series in the \( \theta \) directions 
and take the coefficient of \( \theta_{1}\cdots\theta_{n}, \)
\[ \int d\theta_{1}\cdots d\theta_{n} f(x_{1}, \ldots, x_{k}, 
\theta_{1}, \ldots, \theta_{l}) = f_{i_{1}\cdots i_{n}}(x_{1},\ldots, 
x_{k}). \]

As an application of super geometry we prove the following trivial, 
but crucial, isomorphism.
\begin{lemma}
   \label{lemma:functionsforms}
   Let \( X \) be an even manifold. Then \( C^{\infty}(\Pi 
TX)\cong    \formsij{*}{X} \).  Furthermore, there is a natural 
operator \( Q=\sum_{i}\theta^{i}\partial_{x^{i}}\in C^{\infty}(\Pi TX) 
\) and under the isomorphism we have \( Q\cong d. \)
\end{lemma}
\proof Let \( (x^1,\ldots,x^n) \) be coordinates in a patch on \( X 
\).  Let \( (\theta^1, \ldots, \theta^n) \) be the induced 
coordinates in the odd \( \partial/\partial x^1, \ldots, 
\partial/\partial x^n \) directions of \( \Pi TX \).  Then the isomorphism is simply 
given by 
\[ \theta^i\mapsto dx^i. \]
\subsection{Super Euclidean space}
Consider two 2-dimensional complex vector spaces \spl\ and \sm.  From now on, we shall make use of the notation \( \pm \) to make pairs of statements or definitions at once.  Let
\[ \eps^{\pm}:\wedge^{2}\spm\to \CC \] be fixed isomorphisms.  These 
maps have adjoints
\[ \ad(\eps^{\pm}): \spm\to (\spm)^{*} \]
given by
\[ \ad(\eps^{\pm})(s) = \eps^{\pm}(s, \cdot). \]
The adjoint can be used to define the dual map 
\[ (\eps^{\pm})^{*}:\wedge^{2}(\spm)^{*}\to\CC \]
by mapping

\[ (s_{1}^{\pm}, s_{2}^{\pm})\mapsto 
(s_{1}^{\pm}, \ad(\eps^{\pm})^{-1}(s_{2}^{\pm})) \mapsto 
s_{1}^{\pm} (\ad(\epsilon^{\pm})^{-1}(s_{2}^{\pm})).
\]

Now we choose a basis \( \{\ei{1}, \ei{2}\} \) of \spl, such that \(
\eps^{+}(\ei{1}, \ei{2}) = 1.  \) We also choose a basis
\( \{\eib{1}, \eib{2}\} \) of \sm\ such that \( \eps^{-}(\eib{1},
\eib{2}) = 1.  \) We denote the dual basis by \( \eil{i} \) and \( 
\eilb{i} \). One computes that 
\begin{align}
    \ad(\eps^{\pm})(\eipm{1}) &= \eilpm{2}\label{eq:epslower} \\
    \ad(\eps^{\pm})(\eipm{2}) &= -\eilpm{1}\label{eq:epslower2}
\end{align}
Also, for completeness we have
\begin{align}
    \ad(\eps^{\pm})^{-1}(\eilpm{1}) &= -\eipm{2}\label{eq:epsraise} \\
    \ad(\eps^{\pm})^{-1}(\eilpm{2}) &= \eipm{1}\label{eq:epsraise2}
\end{align}
A final easy computation shows that with the above definition,
\( (\eps^{\pm})^{*}(e^{1}_{\pm}, e^{2}_{\pm}) = 1. \)

We will build a four-dimensional complex vector space \( V_{\CC} \)
with special properties.  First of all, this space has two possible
real structures, which we can use to construct Minkowski space and
Euclidean space.  Second of all, the action of \( V_{\CC} \) on \spm\ 
by Clifford multiplication is ``included'' in the structure of \(
V_{\CC} \) itself.  We'll see more of that shortly.  

We define 
\[ V_{\CC} = \spd\otimes\smd, \]
and now our backward convention of denoting basis vectors with upper 
indices and dual vectors with lower indices should seem justified: \( 
V_{\CC} \) is built from dual spinors.  Thus, elements of \( V_{\CC} 
\) 
have lower indices as expected and only the spinor spaces themselves 
have reversed index conventions.  We equip \( V_{\CC} \) with the 
metric 
\[ \ip{}{} = \frac{1}{2}(\eps^{+})^{*}\otimes(\eps^{-})^{*} \]
or in other words
\[ \ip{s_{1}^{+}\otimes s_{1}^{-}}{s_{2}^{+}\otimes s_{2}^{-}} = 
(\eps^{+})^{*}(s_{1}^{+}, s_{2}^{+})\cdot
(\eps^{-})^{*}(s_{1}^{-}, s_{2}^{-}). \]

To move towards defining a real subspace of \( V_{\CC} \) we note 
that 
we can define maps on spaces with the opposite complex structure,
\[ \opp{(\eps^{\pm})}:\opp{\wedge^{2}\spm}\to \CC, \]
by letting \eps\ act normally on two elements, but then taking the 
complex conjugate of the result:
\[ \opp{(\eps^{\pm})}(a, b) = \overline{\eps^{\pm}(a, b)}. \]

Now we introduce hermitian inner products \( h^{\pm} \) on \spm, which
give isomorphisms
\[ h^{\pm}: \spm\to \opp{(\spm)^{*}}. \]
We require that \( h^{\pm} \) preserve the \eps\ tensors, and so we 
can choose the bases \( e^{i}_{\pm} \) to be orthonormal.

We now use \( h^{\pm} \) to define a real structure on \( V_{\CC} \). 
Consider the map
\[ \ad(\eps^{+})^{-1} \otimes \ad(\eps^{-})^{-1}: 
\spd\otimes\smd\to \spl\otimes\sm \]
and compose it with the map
\[ h^{+}\otimes h^{-}: \spl\otimes \sm\to \opp{\spd}\otimes 
\opp{\smd}. \]
Call this composition \( \tau \).  One easily sees that
\( \tau \) is anti-\CC-linear and that
\( \tau^{2} 
\) is the identity.  We define \( V\subset V_{\CC} \) to be the set 
of 
fixed points of \( \tau \). Below we will see that \( V \) with the metric \ip{}{} is in fact Euclidean 4-space, \( E^4 \). 

It is appropriate to mention the variation of the above construction
that leads to Minkowski space.  First, we begin by setting \(
\sm=\opp{(\spl)}.  \) In other words, \sm\ and \spl\ are opposite
representations of \( SL(2,\CC).  \) We define \( \eps^{-} =
\opp{(\eps^{+})} \), i.e. \( \eps^{-}(a,b) = \overline{\eps^{+}(a,b)}
\).  We define \( \tau \) to be the anti-\CC-linear map exchanging
\sm\ and \spl, which is the identity on the underlying vector spaces,
but which reverses the complex structure.  The fixed set of \( \tau \)
is Minkowski space \(M^4\), with the indefinite metric 
of signature \( (3,1). \)

Let us discuss a permanent change of notation.  Instead of denoting 
elements of \sm\ with minus subscripts, we will place bars over them 
and place dots over their indices.  Therefore, in the new notation a 
basis for \sm\ is denoted \( \{\bar{e}^{\dot{1}}, 
\bar{e}^{\dot{2}}\}. 
\) The dual elements have lower indices.  
Note that the bars and dots do \emph{not} indicate complex 
conjugate.  This awkward-seeming notation is useful to make contact 
with the published physics literature, where Minkowski space is 
usually the context, and as we just saw the elements of \sm\ are the 
conjugates of corresponding elements from \spl. 

Because we are translating 
parts of the physics literature into mathematics, we will have to 
make 
extensive use of index notation.  So we need abbreviations for 
certain 
frequent notation.  For example, we will denote the induced basis on 
the space \( \spl\otimes\spl \) by the four elements \( \euu{a}{b} = 
\ei{a}\otimes \ei{b}. \) A basis of the space \( \spd\otimes\spl \) 
is 
given by elements \( \eul{a}{b} = \eil{b}\otimes \ei{a}. \)  As a 
final 
example, a basis of the space \( \spd\otimes\smd\cong V_{\CC} \) is 
given by \( \eldl{a}{b} = \eil{a}\otimes\eilb{b}. \)

\begin{lemma}
    \( \ip{}{}:V\otimes V\to \CC \) is real and positive definite.
\end{lemma}
\proof Define the following basis of \( V_{\CC} = \spd\otimes\smd \):
\begin{eqnarray}
    v_{1} & = & \eldl{1}{1} + \eldl{2}{2}
    \nonumber  \\
    v_{2} & = & i\eldl{1}{1} - i\eldl{2}{2}
    \label{eq:euclbasis}  \\
    v_{3} & = & \eldl{1}{2} - \eldl{2}{1}
    \nonumber  \\
    v_{4} & = & i\eldl{1}{2} + i\eldl{2}{1}.
    \nonumber 
\end{eqnarray}
Direct computation shows that this basis is real, and that in this basis \ip{}{} is the identity 
matrix.

\begin{definition}
    \( \rrij{4}{4} \) is the subspace \(  
    V\times\Pi(\spd\oplus\smd)\subset V_{\CC}\times\Pi(\spd\oplus\smd). 
    \)
\end{definition}
Note that the base \( V \) is real, while the fibers do not have a 
real structure. 

The automorphism group of \( (S^{\pm}, \eps^{\pm}, h^{\pm}) \) is
\sut, and so \( \sut\times\sut \) acts by isometries on \( V_{\CC}. 
\) This action leaves \( \tau \) invariant and so preserves \( V \),
identifying \( \sut\times\sut \) with the spin double cover of \sof.

\subsubsection{Clifford multiplication}

The special definition of \( V_{\CC} \) makes describing Clifford 
multiplication particularly easy.  The action
\[ V_{\CC}\otimes \spl\to \sm \]
is given by
\[ \spd\otimes\smd\otimes\spl \overset{\mathrm{ev}}{\to}
 \smd \overset{\ad(\eps^-)^{-1}}{\longrightarrow} \sm, \]
where the first map is evaluation on the \spl\ factors. The action of 
\( V_{\CC} \) on \sm\ is given by a similar 
composition
\[ \spd\otimes\smd\otimes\sm \to \spd \to \spl. \]
These actions induce an action of the whole Clifford algebra \( 
Cl(V_{\CC}) \) on \( \spl\oplus\sm \), as one can easily check.  This 
boils down to checking that acting with \( v \) twice 
gives multiplication by \( -\|v\|^{2}. \)

\begin{lemma}
    \label{lemma:stov}
    Clifford multiplication induces isomorphisms
    \begin{align*}
        V_{\CC} &\cong \smd\otimes\spl \\
        \CC\oplus\wedge^{2}_{+}V_{\CC} &\cong \spd\otimes\spl.
    \end{align*}
\end{lemma}

\proof The first isomorphism is given explicitly by
\begin{equation}
    \label{eq:vtohom}
    \tighten{
    \begin{array}{rcrcl}
        v_{1} & \mapsto & -\eudl{2}{1} & + & \eudl{1}{2} \\
        v_{2} & \mapsto & -i\eudl{2}{1} & - & i\eudl{1}{2} \\
        v_{3} & \mapsto & -\eudl{2}{2} & - & \eudl{1}{1} \\
        v_{4} & \mapsto & -i\eudl{2}{2} & + & i\eudl{1}{1} \\
    \end{array}}
\end{equation}
which uses the definition of the \( v_{i} \) together with 
\eqref{epsraise} and \eqref{epsraise2}. 

To prove the second isomorphism we compute as follows.  Compute 
multiplication by \( v_{1}\cdot v_{2} \) (meaning multiply by \( 
v_{2} 
\) and then multiply the result by \( v_{1} \)) with
\begin{equation*}
\begin{split}
    v_{1}\cdot v_{2}\cdot \ei{m} &= v_{1}\cdot 
         \ad(\eps^{-})^{-1}((i\eldl{1}{1} - 
           i\eldl{2}{2})(\ei{m})) \\
    &= v_{1}\cdot \ad(\eps^{-})^{-1}(i\delta^{m}_{1}\eilb{1}
                          -i\delta^{m}_{2}\eilb{2})\\
    &= v_{1}\cdot (-i\delta^{m}_{1}\eib{2}
                          -i\delta^{m}_{2}\eib{1})\\
    &= \ad(\eps^{+})^{-1}((\eldl{1}{1} + \eldl{2}{2})
         (-i\delta^{m}_{1}\eib{2}
                          -i\delta^{m}_{2}\eib{1})) \\
    &= \ad(\eps^{+})^{-1}(-i\delta^{m}_{1}\eil{2} 
                          -i\delta^{m}_{2}\eil{1}) \\
    &= -i\delta^{m}_{1}\ei{1}+i\delta^{m}_{2}\ei{2} \\
    &= i(-1)^m\ei{m}.
\end{split}
\end{equation*}
Doing the calculation for \( v_{3}\cdot v_{4} \) yields an identical 
result, and so using the relationship between wedge product and 
Clifford product, we have computed that 
\[ v_{1}\wedge v_{2} + v_{3}\wedge v_{4}
\mapsto \matrixijkl{-i}{0}{0}{i}. \]
One similarly obtains the rest of the maps
\tighten{
\begin{eqnarray*}
    (1, 0) & \mapsto & \matrixijkl{1}{0}{0}{1} \\
    (0, v_{1}\wedge v_{2} + v_{3}\wedge v_{4}) & \mapsto & 
    \matrixijkl{-i}{0}{0}{i} \\
    (0, v_{1}\wedge v_{3} - v_{2}\wedge v_{4}) & \mapsto & 
    \matrixijkl{0}{1}{-1}{0} \\
    (0, v_{1}\wedge v_{4} + v_{2}\wedge v_{3}) & \mapsto & 
    \matrixijkl{0}{-i}{-i}{0}.
\end{eqnarray*}}%
Noting that \( \mathrm{Hom}(\spl,\spl) 
\cong \spd\otimes\spl \), we represent this as
\begin{equation}
    \label{eq:wedgetohom}
    \tighten{
    \begin{array}{rcrcl}
        (1, 0) & \mapsto & 
        \eul{1}{1} & + & \eul{2}{2} \\
        (0, v_{1}\wedge v_{2} + v_{3}\wedge v_{4}) & \mapsto & 
        -i\eul{1}{1} & + & i\eul{2}{2} \\
        (0, v_{1}\wedge v_{3} - v_{2}\wedge v_{4}) & \mapsto & 
        \eul{1}{2} & - & \eul{2}{1} \\
        (0, v_{1}\wedge v_{4} + v_{2}\wedge v_{3}) & \mapsto & 
        -i\eul{1}{2} & - & i\eul{2}{1}.
    \end{array}}
 \end{equation}
The second isomorphism is now clear. This completes the proof. 

The spaces \( V_{\CC} \) and \( \CC\oplus\wedge^{2}_{+}V_{\CC} \) have 
obvious real subrepresentations, a fact that will be important when 
we twist.

For later use, we provide a version of \eqref{wedgetohom} with 
lowered indices, using \eqref{epslower}, \eqref{epslower2} 
(the raised index becomes the first lower index since \spd\ is the 
first factor).
\begin{equation}
    \label{eq:wedgetohomlower}
    \tighten{
    \begin{array}{rcrcl}
        (1, 0) & \mapsto & 
        \ell{2}{1} & - & \ell{1}{2} \\
        (0, v_{1}\wedge v_{2} + v_{3}\wedge v_{4}) & \mapsto & 
        -i\ell{2}{1} & - & i\ell{1}{2} \\
        (0, v_{1}\wedge v_{3} - v_{2}\wedge v_{4}) & \mapsto & 
        \ell{2}{2} & + & \ell{1}{1} \\
        (0, v_{1}\wedge v_{4} + v_{2}\wedge v_{3}) & \mapsto & 
        -i\ell{2}{2} & + & i\ell{1}{1}.
    \end{array}}
 \end{equation}
 
\subsubsection{Invariant vector fields}
The remainder of this section follows \susols.

We define a coordinate system on \rrij{4}{4} as follows.  Using the 
orthonormal basis elements of \spl, \sm, and \( V \) given above we 
define coordinate functions \( \theta^{a} \) and \( \cjsup{\theta}{a} 
\) on \spd\ and \smd\ respectively (\( a \) and \( \dot{a} \) take on 
values 1 or 2).  On \( V \) we use coordinates that we denote by \( 
y^{a\dot{b}} \) (again, each index takes on the values 1 or 2). So we 
have explicitly
\begin{align}
    \theta^{a}(e_{b}) &= \delta^{a}_{b} \\
    \cjsup{\theta}{a}(\cjsub{e}{b}) &= \delta^{\dot{a}}_{\dot{b}} \\
    y^{a\dot{b}}(e_{c\dot{d}}) &= 
\delta^{a}_{c}\delta^{\dot{b}}_{\dot{d}}.
\end{align}

We denote differentiation in the \( \theta^{a} \) and \(
\cjsup{\theta}{a} \) direction by \di{a} and \dicj{a} respectively. 
We denote differentiation in the \even{y}{a}{b} direction by
\dieven{a}{b}.  We define
\begin{eqnarray*}
    \oda & = & \di{a}- \cjsup{\theta}{b}\dieven{a}{b}\\
    \odacj & = & \dicj{a}- \theta^{b}\dieven{b}{a}
    \label{eq:defoda}
\end{eqnarray*}
and
\begin{eqnarray*}
    \oqa   & = & \di{a}+ \cjsup{\theta}{b}\dieven{a}{b}\\
    \oqacj & = & \dicj{a}+ \theta^{b}\dieven{b}{a}.
    \label{eq:defoqa}
\end{eqnarray*}
These vector fields satisfy the bracket relations (remembering that 
for two odd vector fields you add instead of subtract to form 
brackets)
\begin{align*}
[\oda,\odi{b}]    & = [\odacj,\odicj{b}]=0\\%
{}[\oda,\odicj{b}]  & = -2\dieven{a}{b}
\end{align*}
and
\begin{align*}
[\oqa, \oqi{b}]   & =  [\oqacj,\oqicj{b}] = 0 \\%
{}[\oqa, \oqicj{b}] & = 2\dieven{a}{b}.
\end{align*}

\subsubsection{\mytexorpdfstring{\rrij{4}{8}}{E(4|8)}}

Simply put, \( N=2 \) super Euclidean space, also called \rrij{4}{8},
has two copies of \( \Pi(\spd\oplus\smd) \) instead of one.  We don't
need to repeat the above discussion, but there are some 
complications. 
First of all, we need to provide the odd coordinate functions and
vector fields with another index, that can take on the values 1 or 2,
to represent which copy of \( \Pi(\spd\oplus\smd) \) they live on. 
So, we now have the odd coordinate functions 
\[ \theta^{1(1)},\theta^{2(1)},
\overline{\theta}^{\dot{1}(1)},
\overline{\theta}^{\dot{2}(1)},
\theta^{1(2)},\theta^{2(2)},
\overline{\theta}^{\dot{1}(2)},
\overline{\theta}^{\dot{2}(2)} \] 
as well as the left-invariant vector fields 
\begin{gather*}
\one{\odi{1}}, \one{\odi{2}},
\one{\odicj{1}}, \one{\odicj{2}} \two{\odi{1}}, \two{\odi{2}},
\two{\odicj{1}}, \two{\odicj{2}}, \\
 \one{\oqi{1}}, \one{\oqi{2}},
\one{\oqicj{1}}, \one{\oqicj{2}} \two{\oqi{1}}, \two{\oqi{2}},
\two{\oqicj{1}}, \two{\oqicj{2}}.
\end{gather*}
The commutation relations are the same as before, with brackets of
vector fields of differing upper index vanishing: \tighten{
\begin{eqnarray}
 [\onetwoi{\oda},\onetwoij{\odi{b}}{j}]    & = & 
 [\onetwoi{\odacj},\onetwoij{\odicj{b}}{j}]=0   \nonumber \\%
 {}[\onetwoi{\oda},\onetwoij{\odicj{b}}{j}]  & = & 
-2\delta^{ij}\dieven{a}{b} \label{eq:e48dbracket}
\end{eqnarray}}%
and
\tighten{
\begin{eqnarray}
 [\onetwoi{\oqa},\onetwoij{\oqi{b}}{j}]    & = & 
 [\onetwoi{\oqacj},\onetwoij{\oqicj{b}}{j}]=0  \nonumber  \\%
 {}[\onetwoi{\oqa},\onetwoij{\oqicj{b}}{j}]  & = & 
2\delta^{ij}\dieven{a}{b}. \label{eq:e48qbracket}
\end{eqnarray}}%
To sum up the index structure of these vector fields we make the 
following remark.
\begin{obs}
    The \( Q \)'s and \( D \)'s are sections of the \( 
\spin(4)\times 
    \sut\)-bundle \( \spd\otimes\CC^{2} \) 
    and the \( \bar{Q} \)'s and \(\bar{D} \)'s are sections of the \( 
    \spin(4)\times \sut\)-bundle \( \smd\otimes\CC^{2} \). 
    \label{obs:qsassections}
\end{obs}

\subsection{Gauge theory on \mytexorpdfstring{$\rrij{4}{8}$}{E(4|8)}}

Much of this section can be considered ``standard material'' and can 
be found in the literature.  One thorough account can be found in 
\cite{phong}.  Another good accounting, and the one whose notation we 
adopt here, is \susols.

Deciding what category of connections we should work with is a subtle 
business.  The correct formulation in \( N=1 \) theories from the 
physical standpoint is to examine \emph{constrained} connections.

\begin{definition}
    A superconnection \aa\ on a supermanifold with odd 
    distribution \( \tau \)
    is said to be \emph{constrained} if the curvature \ff\ of \aa\ 
    vanishes along \( \tau \).  That is, if \( \ff(x, y) = 0 \) 
    whenever \( x \) and \( y \) are odd vector fields.
\end{definition}
There are physical reasons for requiring this, but from the
mathematical perspective it's just a subcategory we happen to be
focusing on.  Things are different in \( N=2 \) theories, though. 
Here there are eight odd directions to consider in four-dimensional
theories.  One proceeds by considering \( N=1 \) superconnections in
\emph{six} dimensions, where the spin bundle has eight
dimensions.  We will reduce this picture to four dimensions by
requiring translation invariance along two dimensions, say the span of
\( v \) and \( w \) for \( v, w\in\RR^{6} \) linearly independent.  So
we examine a principal \sut\ bundle \pp\ over \rrij{6}{8} that is 
trivial in
the \( v \) and \( w \) directions.  Then we work with constrained
connections that are constant along \( v \) and \( w \).  

We will see in a moment that such a dimensionally reduced object is 
no longer constrained.  Instead, it can have two independent 
nonvanishing scalar curvatures on the odd distribution of 
\rrij{4}{8}.  

\begin{definition}
    A superconnection on a supermanifold whose curvature
    vanishes identically along the odd distribution except for two 
    two-dimensional subdistributions along which the curvature is 
    unconstrained is called \emph{semi-constrained}.
\end{definition}

\begin{theorem}
    \label{thm:semiconstrained}
    The space of dimensionally reduced connections from \rrij{6}{8} to
    \rrij{4}{8} is isomorphic to the space of semi-constrained
    connections on \rrij{4}{8}.
\end{theorem}

\proof 
It is easiest to use proper coordinates and vector fields from 
\rrij{6}{8}, so we give a brief run-down of this.  Details can be 
found in \susols.
We will use the name \( y^{ab} \) and \( \theta^{ai} \) for the 
coordinate system on \rrij{6}{8}, and \( \di{ab} \) and \( 
\di{ai} \) for the corresponding vector fields.  Here, \( a, b \) 
take 
on the values 1 through 4, but with \( a<b \).  \( i \) can be 1 or 
2. We denote the \(\epsilon\) tensor in coordinates as 
\(\epsilon_{ij}\) (\( i \) and \( j \) can take on values 1 or 2).

The eight left-invariant vector fields are given by
\begin{equation}
    \odi{ai} = \di{ai} - \epsilon_{ij}\theta^{bj}\di{ab}
    \label{eq:defofdinm68}
\end{equation}
and the right-invariant ones by
\begin{equation}
    \oqi{ai} = \di{ai} + \epsilon_{ij}\theta^{bj}\di{ab}
    \label{eq:defofqinm68}
\end{equation}
with the commutation relations
\begin{eqnarray*}
    \left[\odi{ai}, \odi{bj}\right] & = & -\epsilon_{ij}\di{ab}  \\
    \left[\oqi{ai}, \oqi{bj}\right] & = & +\epsilon_{ij}\di{ab}.
\end{eqnarray*}

By dimensional reduction we mean the restriction to \rrij{4}{8}, 
which is just the standard embedding of \(\RR^4\) into \(\RR^6\) by 
setting two coordinates on \(\RR^6\) to 
zero.  The effect on the coordinate systems we've been using is 
\begin{alignat}{2}
    y^{12} & = 0 
    & \quad y^{23} & = \even{y}{2}{1}\nonumber\\
    y^{13} & = \even{y}{1}{1}
    & \quad y^{24} & = \even{y}{2}{2}\label{eq:dimred2}\\
    y^{14} & = \even{y}{1}{2}
    & \quad y^{34} & = 0.\nonumber
\end{alignat}

We do not reduce the number of odd coordinates, though, and we can 
make a dictionary of left-invariant vector fields
\begin{alignat}{2}
    D_{11} & = \onetwoij{\odi{1}}{1}
    & \quad D_{12} & = \onetwoij{\odi{1}}{2}\nonumber\\
    D_{21} & = \onetwoij{\odi{2}}{1}
    & \quad D_{22} & = \onetwoij{\odi{2}}{2} \label{eq:dimred}\\
    D_{31} & = -\onetwoij{\odicj{1}}{2}
    & \quad D_{32} & = \onetwoij{\odicj{1}}{1}\nonumber\\
    D_{41} & = -\onetwoij{\odicj{2}}{2}
    & \quad D_{42} & = \onetwoij{\odicj{2}}{1}\nonumber.
\end{alignat}

We are reducing by two 
dimensions, from \rrij{6}{8} to \rrij{4}{8}, by setting the 
coordinates \( y^{12} \) and \( y^{34} \) to zero.  Whereas we have
on \rrij{6}{8} the relation
\begin{equation}
     [\odi{31}, \odi{42}]=-\di{34},
    \label{eq:m68bracket}
\end{equation}
under the reduction correspondence \eqref{dimred}, \( \odi{31} =
-\onetwoij{\odicj{1}}{2} \), \( \odi{42} = \onetwoij{\odicj{2}}{1}, \)
we instead have on \rrij{4}{8} the equation
\begin{equation}
    -[\onetwoij{\odicj{1}}{2}, \onetwoij{\odicj{2}}{1}] = 0.
    \label{eq:m48bracket}
\end{equation}

The covariant version of \eqref{m68bracket} is
\begin{equation}
     [\odicov{31}, \odicov{42}]=-\dicov{34},
    \label{eq:m68bracketcov}
\end{equation}
and we can wonder, What happens to this equation after we
dimensionally reduce?  The two odd vector fields become two of the
vector fields on \rrij{4}{8}, but \di{34} becomes zero, so what is the
reduction of this covariant equation?  The answer is that if the
principal bundle \pp\ and the connection are invariant under
translations in the \( y^{34} \) direction, then there is a trivial
lift of \( \di{34} \) which we will call \ditil{34}; it is the lift of
\di{34} using the product connection in this trivial direction.  The
difference \( \dicov{34} - \ditil{34} \) is a vertical vector field,
or a section of the adjoint bundle.  Dimensional reduction simply
states that there can be no component of the lift of \(
[\onetwoij{\odicj{1}}{2}, \onetwoij{\odicj{2}}{1}] \) in the
\ditil{34} direction, and so this bracket must lift to the
\emph{vertical part} of \( -\dicov{34} \).  We define 
\[ \Sigma =%
\dicov{34} - \ditil{34} \] and thus have the
dimensionally reduced equation
\begin{equation}
    [\onetwoij{\odicjcov{1}}{2}, \onetwoij{\odicjcov{2}}{1}] = 
    \Sigma.
    \label{eq:sigma1}
\end{equation}

This equation tells us that this particular component of odd-odd 
curvature \emph{need not vanish}.  

Similarly, we have
\[ [\odi{41}, \odi{32}] = \di{34} \]
so using the correspondence \( \odi{41} = -\onetwoij{\odicj{2}}{2} 
\), 
\( \odi{32} = \onetwoij{\odicj{1}}{1}  \) we get
\begin{equation}
    -[\onetwoij{\odicj{2}}{2}, \onetwoij{\odicj{1}}{1}] = \Sigma.
    \label{eq:sigma2}
\end{equation}

Now, we consider constancy in the \( y^{12} \) direction.  This leads 
to a second section of the adjoint bundle that we'll call \( 
\bar{\Sigma} \)
\[ \bar{\Sigma} = \dicov{12} - \ditil{12}. \]
This in turn leads to the equations
\begin{eqnarray}
    -\left[\onetwoij{\odicov{2}}{2}, \onetwoij{\odicov{1}}{1}\right] 
& = &  
    \bar{\Sigma} \label{eq:sigma3} \\
    \left[\onetwoij{\odicov{1}}{2}, \onetwoij{\odicov{2}}{1}\right] & 
= &  
    \bar{\Sigma} \label{eq:sigma4}.
\end{eqnarray}
This completes the proof.

We will see this theorem play out in the twisted context as well, 
where we will have two independent odd-odd curvatures that are not 
required to vanish.

\subsubsection{Component Fields}
The component fields of a superconnection in \rrij{4}{8} are denoted 
\( A, \sigma, \overline{\sigma}, 
\lambda, \overline{\lambda}, \chi, 
\overline{\chi}, E, F. \) These are
\begin{equation}
    \begin{split}
        A                  &= \text{a connection on } E^{4} \\
        \sigma             &= \text{a section of }  \adp  \\
        \overline{\sigma}  &= \text{a section of }  \adp  \\
        \lambda            &= \text{a section of }  \adp\otimes\spd  \\
        \overline{\lambda} &= \text{a section of }  \adp\otimes\smd  \\
        \chi               &= \text{a section of } \adp\otimes\spd  \\
        \overline{\chi}    &= \text{a section of }  \adp\otimes\smd  \\
        E                  &= \text{a section of }  \adp  \\
        F                  &= \text{a section of }  \adp\otimes\CC  \\
    \end{split}
    \label{eq:defofcomponentsinr48}
\end{equation}
These are defined as follows.  \( A \) is the induced connection on 
the induced even bundle \( P\to E^{4} \) sitting inside \( 
\pp\to\rrij{4}{8}. \)  The others are defined by
\begin{align}
    \sigma &= i^{*}\Sigma\nonumber \\
    \bar{\sigma} &= i^{*}\bar{\Sigma}\nonumber \\
    \lambda_{a} &= i^{*} W^{1}_{a} = i^{*} 
     \frac{1}{4}\eps^{\dot{c}\dot{d}} [\odijcjcov{1}{c}, 
\dievencov{a}{d}] 
     \nonumber\\
    \chi_{a} &= i^{*} W^{2}_{a} = i^{*} 
     \frac{1}{4}\eps^{\dot{c}\dot{d}} [\odijcjcov{2}{c}, 
     \dievencov{a}{d}]\nonumber \\
    \overline{\lambda}_{\dot{a}} &= i^{*} \overline{W}^{1}_{\dot{a}} 
= i^{*} 
     \frac{1}{4}\eps^{cd} [\odijcov{1}{c}, 
     \dievencov{d}{a}]\label{eq:defofcomponentsinr482} \\
    \overline{\chi}_{\dot{a}} &= i^{*} \overline{W}^{2}_{\dot{a}} = 
i^{*} 
     \frac{1}{4}\eps^{cd} [\odijcov{2}{c}, \dievencov{d}{a}] 
\nonumber\\
    E &=  -i^{*}(\one{\odicjcov{2}} 
     \two{\odicjcov{1}}\Sigma - \one{\odicjcov{1}} 
     \two{\odicjcov{2}}\Sigma)\nonumber \\
    F &= i^{*}\two{\odicjcov{2}}\two{\odicjcov{1}}
     \Sigma\nonumber \\
    \bar{F} &= i^{*}\one{\odicjcov{1}}\one{\odicjcov{2}}
     \Sigma\nonumber.
 \end{align}
Here, \( i^* \) is the pullback functor using the inclusion \( 
i:E^4\hookrightarrow \rrij{4}{8}. \)  

\begin{theorem}
    The space of semi-constrained superconnections on \rrij{4}{8} is 
    isomorphic to the space of component fields.
\end{theorem}
See \susols\ for a discussion of this.

\subsection{The super Yang-Mills \action}
\label{flataction}
Let \( \tau \) be the complex parameter
\begin{equation}
    \label{eq:tau}
    \tau = \frac{\theta}{2\pi} + \frac{4\pi i}{g^{2}}.
\end{equation}
The \action\ on super Minkowski space \mij{4}{8} is given by
\begin{equation}
    S = \int d^{4}x\, \im\left( 
    d^{4}\theta\,\frac{\tau}{32\pi}
    \langle\Sigma, \Sigma\rangle
    \right)
    \label{eq:superaction}
\end{equation}
We will write this action in its component
formulation.  This is obtained from \eqref{superaction} by integrating
out the four odd variables, or equivalently, hitting the integrand
with an appropriate combination of four odd derivatives.  In this case
that is 
\begin{equation}
 \label{eq:fouroddder}
 \one{\odi{2}}\one{\odi{1}}\two{\odi{2}}\two{\odi{1}},
\end{equation} 
though other choices are appropriate as well, so long as they differ 
from this one by an exact term.
See \susols\ for more information about this computation.  The Dirac
pairing seen below is defined by
\begin{equation}
    \label{eq:diracpairing}
    \langle{\lambda}\odiraca_{A}\bar{\lambda}\rangle = 
    \epsilon^{ac}\epsilon^{\dot{b}\dot{d}}
    \lambda_{c}\nabla_{a\dot{b}}\bar{\lambda}_{\dot{d}}.
\end{equation}
The action in components is therefore given by    
\tighten{
\begin{eqnarray}
    S  = \int d^{4}x\,\frac{1}{g^{2}}\Big\{ & - & 
    \frac{1}{2}|F_{A}|^{2}
    + \ip{d_{A}\bar{\sigma}}{d_{A}\sigma} 
    + \langle{\lambda}\odiraca_{A}\bar{\lambda}\rangle
    + \langle{\chi}\odiraca_{A}\bar{\chi}\rangle
    \nonumber  \\
    & - & 
     \eps^{ab}\ip{\bar{\sigma}}{[\lambda_{a},\chi_{b}]}
               + \eps^{\dot{a}\dot{b}}
               \ip{[\cjsub{\lambda}{a}, \cjsub{\chi}{b}]}{\sigma}
     - \frac{1}{2} |E|^{2}
    \label{eq:m4action}  \\
    & + & \ip{\bar{F}}{F}\Big\}
     + \frac{\theta}{16\pi^{2}}\langle F_{A}\wedge F_{A}\rangle.
    \nonumber
\end{eqnarray}}
Note the presence of the usual Yang-Mills action (the first term), as
well as the second Chern class (the last term). However, we are 
following the usual convention of having separate coupling 
coefficients for these two terms.  This is because the topological 
Chern-Simons term has a different character in the physical theory, 
since it is locally constant on components of \aa.  We will find 
reason to revisit the value of the coefficient of the ``theta term'' 
when we twist the action in \secref{twistedaction}.

Next we write the result of Wick rotating this action to \rrij{4}{8}.  
This is a procedure we will not carry out explicitly, but merely write 
the result.  For more information, see \cite{wickrotation}.  It 
differs only in the signs and some coefficients of \( i \). 
\tighten{
\begin{eqnarray}
    S  = \int d^{4}x\,\frac{1}{g^{2}}\Big\{ & - & 
      \frac{1}{2}|F_{A}|^{2}
    - i\ip{d_{A}\bar{\sigma}}{d_{A}\sigma} 
    + i\langle{\lambda}\odiraca_{A}\bar{\lambda}\rangle
    + i\langle{\chi}\odiraca_{A}\bar{\chi}\rangle
    \nonumber  \\
    & + & 
    i\eps^{ab}\ip{\bar{\sigma}}{[\lambda_{a},\chi_{b}]}
               + i\eps^{\dot{a}\dot{b}}
               \ip{\sigma}{[\cjsub{\lambda}{a}, \cjsub{\chi}{b}]}
     - i\frac{1}{2} |E|^{2}
    \label{eq:e4action}  \\
    & + & i\ip{\bar{F}}{F}\Big\}
     + \frac{\theta}{16\pi^{2}}\langle F_{A}\wedge F_{A}\rangle.
    \nonumber
\end{eqnarray}}

The material in this section would greatly benefit from a treatment more within 
the philosophical scope of this paper. This remark applies equally to the 
computation of the twisted counterpart to the above action formula, which is a 
main result of the next section. The enterprising reader should perhaps focus 
on the operator \eqref{fouroddder}, in search of a generalized interpretation 
and formula for an operator that integrates over the odd components of a 
superspace. 
\section{The superspace \mytexorpdfstring{$SX$}{SX}}

\label{sxchapter}

\subsection{The twist}
The twisting operation is a modification of the global structure of 
\rrij{4}{8}.  It will alter the structure of any theory over this 
base, and so we will obtain a class of theories that is very different 
from those on honest supersymmetric space.  However, this trade-off 
allows us to construct an extension to any 4-manifold \( X \) that is 
analogous to extending \( E^{4} \) to \rrij{4}{8}.

The presence of the \( \delta \)-function on the right hand side of
the bracket relations \eqref{e48dbracket} and \eqref{e48qbracket} is a
clue that there is an automorphism group of \rrij{4}{8} that preserves
the even subspace.  We can see immediately that the group that acts on
the \onetwoi{\oda} and \onetwoi{\odacj} preserving the bracket is the
group \( U(2) \) preserving a symmetric hermitian bilinear pairing on
\( \CC^{2} \).  One calls this \( U(2) \) the \emph{\( R \)-symmetry
group}.  We will only be discussing the subgroup \( \sut\subset U(2). \)
The quotient group \( U(2)/\sut\cong U(1) \) plays quite a different
role in the physics, and presumably in the mathematics as well, related 
to anomalies, but we will not encounter it further.  We denote the \sut\ 
part of the \(R\)-symmetry by \(\sut^R\).

The odd fiber of \rrij{4}{8} is a representation of \(\spin(4)\) as well as of 
\(\sut^R\).
We denote the decomposition of \( \spin(4) \) as \( \spin(4)\cong 
\sut^+\times\sut^- \).
The \( R \)-symmetry
allows us to construct an interesting and important map \( \spin(4) \) into \( H =
\sut^{+}\times\sut^{-} \times\sut^{R} \). We define the \emph{twist 
homomorphism}
\[ T:\sut^{+}\times\sut^{-} \to H \]
by
\begin{equation}
    T(a, b) = (a, b, a)
    \label{eq:twist}
\end{equation}
which is the diagonal embedding of \sut\ into \( 
\sut^{+}\times\sut^{R} 
\) combined with the identity mapping of \( \sut^{-} \).  
Clearly \(H\) acts on \( \CC^2\otimes\CC^2\otimes\CC^2 \), so we can now form a new \( \spin(4) \) associated vector bundle over \(E^4\) with fiber \( \CC^2\otimes\CC^2\otimes\CC^2 \) by precomposing with the mapping \( T \).
Another way 
to look at this operation is that we have declared that the index for 
the trivial \( \CC^{2} \) fiber of \rrij{4}{8} now labels another copy of \spl\ instead.

We can use this to do something very special.  We can use the 
isomorphisms in \lemmaref{stov} to
prove immediately that these bundles factor through \sof, and so can 
be formed on any riemannian 4-manifold.  Since the twisted
vector fields take values in \( \spd\otimes\spl \) and \(
\smd\otimes\spl \), then after the twist
one vector field takes values in \(
\RR \), one of them takes values in \( \wedge^{2}_{+} V \), and one of
them takes values in \( V \).  Explicitly we define
\begin{eqnarray}
    D_{0} & = & \dul{1}{1} + \dul{2}{2}\label{eq:d0} \\
    &&\nonumber\\
    D_{1} & = & \left(\dudl{1}{2} - \dudl{2}{1}\right) dv^{1}+
    \left(- i\dudl{1}{2} - i\dudl{2}{1}\right) dv^{2}\label{eq:d1}\\
     & + & \left(-\dudl{1}{1} - \dudl{2}{2} \right) dv^{3}+
     \left(i\dudl{1}{1} - i\dudl{2}{2}\right) dv^{4}\nonumber\\
    &&\nonumber\\
    D_{2} & = & \left(i\dul{1}{1} + i\dul{2}{2}\right) 
                (dv^{1}\wedge dv^{2} + dv^{3}\wedge dv^{4})\nonumber\\
    & + & \left(\dul{1}{2} - \dul{2}{1}\right) 
                (dv^{1}\wedge dv^{3} - dv^{2}\wedge 
dv^{4})\label{eq:d2}\\
    & + & \left(-i\dul{1}{2} - i\dul{2}{1}\right) 
                (dv^{1}\wedge dv^{4} + dv^{2}\wedge dv^{3})\nonumber.
\end{eqnarray}

These three vector fields should have a more intrinsic description, 
one that does
not make reference to supersymmetry or the spin bundles we have
tensored together.  The results along these lines are as follows.

Let \( X \) be a riemannian four-manifold with local coordinate
functions \( x^{i} \).  Form the odd vector bundle 
\[ SX = \Pi\left(
(X\times\CC)\oplus TX\oplus \wedge^{2}_{+}TX\right).  \]
The \( x^{i} \) induce local bases \( \partial_{x^{i}} \) for vector
fields, and \( dx^{i} \) for one-forms.  We obtain induced coordinates
in the \( \Pi TX \) directions that we will denote by \( \theta^{i} \)
(so \( \theta^{i} \) is an odd coordinate function along the odd \(
\partial_{x^{i}} \) direction).  Similarly the \( x^{i} \) induce
coordinates in the \( \wedge^{2}_{+}TX \) directions.  We will denote
by \( \theta^{1234} \) the coordinate in the odd \( \wdd{1}{2} +
\wdd{3}{4} \) direction, \( \theta^{1324} \) in the odd \( \wdd{1}{3} 
-
\wdd{2}{4}\) direction, and \( \theta^{1423} \) in the odd \( 
\wdd{1}{4} +
\wdd{2}{3} \) direction.  Lastly, we will call the coordinate in the
trivial odd direction \( \theta \).

Define 
\begin{eqnarray*}
 M & = & \theta^{1234}\otimes(\wdu{1}{2} - \wdu{2}{1} +
 \wdu{3}{4} - \wdu{4}{3}) \\
 & + & \theta^{1324}\otimes(\wdu{1}{3} - \wdu{3}{1} -
 \wdu{2}{4} + \wdu{4}{2}) \\
 & + & \theta^{1423}\otimes(\wdu{1}{4} - \wdu{4}{1} +
 \wdu{2}{3} - \wdu{3}{2}).
\end{eqnarray*}
We can now state intrinsic (although coordinate-dependent) formulas 
for the \( D_{i} \).

\begin{prop} \( D_{0} = \partial_{\theta} - 
\theta^{i}\partial_{x^{i}} 
\), 
\( D_{1} = \partial_{\theta^{i}}dx^{i} - \theta\partial_{x^{i}} 
dx^{i} - M \) and 
\begin{align*}
    D_{2} & = \partial_{\theta^{1234}}
    (dx^{1}\wedge dx^{2}+dx^{3}\wedge dx^{4}) \\
    & + \partial_{\theta^{1324}}
    (dx^{1}\wedge dx^{3}-dx^{2}\wedge dx^{4}) \\
    & + \partial_{\theta^{1423}}
    (dx^{1}\wedge dx^{4}+dx^{2}\wedge dx^{3}) \\
    & - (\theta^{1}\partial_{x^{2}} - \theta^{2}\partial_{x^{1}}
    +\theta^{3}\partial_{x^{4}} - \theta^{4}\partial_{x^{3}})
    (dx^{1}\wedge dx^{2}+dx^{3}\wedge dx^{4}) \\
    & - (\theta^{1}\partial_{x^{3}} - \theta^{3}\partial_{x^{1}}
    -\theta^{2}\partial_{x^{4}} + \theta^{4}\partial_{x^{2}})
    (dx^{1}\wedge dx^{3}-dx^{2}\wedge dx^{4}) \\
    & - (\theta^{1}\partial_{x^{4}} - \theta^{4}\partial_{x^{1}}
    +\theta^{2}\partial_{x^{3}} - \theta^{3}\partial_{x^{2}})
    (dx^{1}\wedge dx^{4}+dx^{2}\wedge dx^{3})
\end{align*}
\end{prop}

\proof First a bit of motivation.  The vector fields 
\dul{a}{b} and \upperlowerabc{\bar{D}}{a}{\dot{b}} are made of two 
terms:  a partial derivative 
in an odd direction and an odd coordinate function times an even 
partial 
derivative.  Let us try to construct global objects with this form.  
Guessing at the formula for \( D_{0} \), for example, is easy if you 
want to obtain 
this form. The others are more complex.

To construct the 1-form \( D_{1} \), we can take advantage of the
redundancy in having \( X \) and \( \Pi TX \) both available, and use
the isomorphism between the odd and even tangent spaces.  This is what
\( \partial_{\theta^{i}}dx^{i} \) does.  We can also construct de Rham
\( d \), the identity element in \( Hom(TX, TX), \) and multiply it by
\( \theta \).  Lastly for \( D_{1} \) we can try to find an element of
\[ \Pi\wedge^{2}_{+}TX \otimes TX\otimes T^{*}X \]
where the first factor are the \( \theta^{abcd} \) coefficients, the 
second are the vector fields, and the third are the \( dx^{i} \).  Is 
there a canonical element of this bundle?  Yes, look at
\[ \mathrm{Id}\in\Pi \wedge^{2}_{+}TX\otimes\wedge^{2}_{+}T^{*}X \]
and map it through the inclusion
\[ \Pi \wedge^{2}_{+}TX\otimes\wedge^{2}_{+}T^{*}X \hookrightarrow 
\Pi\wedge^{2}_{+}TX\otimes T^{*}X\otimes T^{*}X \]
followed by taking the dual on the first \( T^{*}X \) using the 
metric.  This is the element \( M \).

For \( D_{2} \) we can use the isomorphism between the even and the
odd self-dual 2-vectors, which is what the first three terms of \(
D_{2} \) do.  For the second set of three terms we take the element \[
\mathrm{Id}\in\wedge^{2}_{+}TX\otimes\wedge^{2}_{+}T^{*}X \] and map
it through the inclusion \[
\wedge^{2}_{+}TX\otimes\wedge^{2}_{+}T^{*}X\hookrightarrow TX\otimes
TX\otimes\wedge^{2}_{+}T^{*}X \] followed by taking parity reversal on
the first \( TX \).  Although these three constructions are very
canonical and unique, they do not suffice to prove the relationship to
the \rrij{4}{8} formulas.  However, once we have proven this
relationship rigorously, we should leave the Proposition with the
sense that twisted supersymmetry has a very rich and deep relationship
to intrinsic smooth objects.

We will prove the formula for \( D_{0} \) as an example, and leave 
the 
rest as exercises.  \( \theta 
\) is the coordinate in the trivial line bundle direction, and we 
have 
from \eqref{wedgetohom} that
\[ \theta = \theta^{1}_{1} + \theta^{2}_{2}. \]
To write out \( \theta^{i}\partial_{i} \) we use \eqref{euclbasis} 
directly  on 
the partials, and use \eqref{euclbasis} plus the usual change of variables 
formula for the \( \tudu{a}{b} 
\)'s.  The change of variables has the effect of taking the complex 
conjugate compared to 
the \( \partial \) formulas, which is analogous to the relationship 
\( 
z = x+iy, \partial_{z} = \partial_{x} - i\partial_{y}. \)
\begin{align*}
    \theta^{1}\partial_{1} &= ( \tudu{1}{1} +  \tudu{2}{2})
                              ( \pldl{1}{1} +  \pldl{2}{2}) \\
    \theta^{2}\partial_{2} &= (-i\tudu{1}{1} + i\tudu{2}{2})
                              (i\pldl{1}{1} - i\pldl{2}{2}) \\
    \theta^{3}\partial_{3} &= ( \tudu{1}{2} -  \tudu{2}{1})
                              ( \pldl{1}{2} -  \pldl{2}{1}) \\
    \theta^{4}\partial_{4} &= (-i\tudu{1}{2} - i\tudu{2}{1})
                              (i\pldl{1}{2} + i\pldl{2}{1}).
\end{align*}
Adding this all up gives
\[ \partial^{1}_{1} + \tudu{1}{1}\pldl{1}{1} + 
\tudu{1}{2}\pldl{1}{2} + \partial^{2}_{2} + 
\tudu{2}{1}\pldl{2}{1} + \tudu{2}{2}\pldl{2}{2} = \qul{1}{1} + 
\qul{2}{2}, \]
as required.

We also define \( Q \) analogues of \( D_0 \), \( D_1 \) and \( D_2 
\), but with plus signs instead of minus signs. Explicitly we have  
\( Q_{0} = \partial_{\theta} + \theta^{i}\partial_{x^{i}} 
\), 
\( Q_{1} = \partial_{\theta^{i}}dx^{i} + \theta\partial_{x^{i}} 
dx^{i} + M \) and 
\begin{align*}
    Q_{2} & = \partial_{\theta^{1234}}
    (dx^{1}\wedge dx^{2}+dx^{3}\wedge dx^{4}) \\
    & + \partial_{\theta^{1324}}
    (dx^{1}\wedge dx^{3}-dx^{2}\wedge dx^{4}) \\
    & + \partial_{\theta^{1423}}
    (dx^{1}\wedge dx^{4}+dx^{2}\wedge dx^{3}) \\
    & + (\theta^{1}\partial_{x^{2}} - \theta^{2}\partial_{x^{1}}
    +\theta^{3}\partial_{x^{4}} - \theta^{4}\partial_{x^{3}})
    (dx^{1}\wedge dx^{2}+dx^{3}\wedge dx^{4}) \\
    & + (\theta^{1}\partial_{x^{3}} - \theta^{3}\partial_{x^{1}}
    -\theta^{2}\partial_{x^{4}} + \theta^{4}\partial_{x^{2}})
    (dx^{1}\wedge dx^{3}-dx^{2}\wedge dx^{4}) \\
    & + (\theta^{1}\partial_{x^{4}} - \theta^{4}\partial_{x^{1}}
    +\theta^{2}\partial_{x^{3}} - \theta^{3}\partial_{x^{2}})
    (dx^{1}\wedge dx^{4}+dx^{2}\wedge dx^{3}).
\end{align*}
Using these formulas, it is trivial to check that all three \( Q \)'s 
commute with all three \( D \)'s.  Also, other commutators that 
will interest us are
\begin{align}
[D_0, D_1] &= -d \\
[Q_0,Q_1]  &= d.
\end{align}
Here, and from now on, we will use the following definition of the 
bracket
\begin{equation}
\label{eq:newbracket}
[A,B] = \frac{1}{2}(AB - (-1)^{\pi(A)\pi(B)}BA).
\end{equation}

\subsection{Superconnections on \mytexorpdfstring{$SX$}{SX}}

The general idea is that we will simply twist the picture presented 
in 
\eqref{defofcomponentsinr48}.  For 
example, the two \spl-valued sections \( \lambda \) and \( \chi \) of 
\( \adp \) will combine to form a single \( 
\spd\otimes\spl \)-valued section of \( \adp \), and so will 
decompose as a section of \( \adp \) and a section of \( \ad\, 
P\otimes\wedge^{2}_{+}TX \).  

Let us begin with the spinors \( \lambda \) and \( \chi \).  In flat 
space they are defined by 
\begin{gather}
\chi_{a} = i^{*} W^{1}_{a} = i^{*} 
\frac{1}{4}\eps^{\dot{c}\dot{d}} [\odijcjcov{1}{c}, \dievencov{a}{d}] 
\\
\lambda_{a} = i^{*} W^{2}_{a} = i^{*} 
\frac{1}{4}\eps^{\dot{c}\dot{d}} [\odijcjcov{2}{c}, \dievencov{a}{d}].
\end{gather}
These can therefore equivalently be defined as the image of the expression
\[ \frac{1}{4}[\odijcjcov{i}{c}, \dievencov{a}{d}] \]
under the mapping
\[ 
(\eps^{-})^{*}:\smd\otimes\spl\otimes\smd\otimes\spd\to\spl\otimes\spd. 
\]
A computation shows that this operation can be interpreted quite 
simply in global language.  Precisely, we have
\begin{prop}
\label{prop:mapsforw}
    The following diagram commutes
    \begin{equation}
        \begin{CD}
            \smd\otimes\spl\otimes\smd\otimes\spd @>(\eps^{-})^{*}>> 
            \spl\otimes\spd \\
            @VV\cong V            @VV\cong V \\
            TX\otimes TX @>\mathrm{proj.}>> 
(X\times\RR)\oplus\wedge^{2}_{+}TX
        \end{CD}
        \label{eq:mapsforw}
    \end{equation}
\end{prop}

The barred spinors \( \bar{\lambda} \) and \( \bar{\chi} \) are 
defined by
\begin{gather}
    \overline{\lambda}_{\dot{a}} = i^{*} \overline{W}^{1}_{\dot{a}} = 
i^{*} 
    \frac{1}{4}\eps^{cd} [\odijcov{1}{c}, \dievencov{d}{a}] 
    \label{eq:lambda}\\
    \overline{\chi}_{\dot{a}} = i^{*} \overline{W}^{2}_{\dot{a}} = 
i^{*} 
    \frac{1}{4}\eps^{cd} [\odijcov{2}{c}, 
    \dievencov{d}{a}].\label{eq:chi}
\end{gather}
These are the images of the expressions
\[ \frac{1}{4}[\odijcov{i}{c}, \dievencov{d}{a}] \]
under the mapping
\[ \epsilon^{+}: \spd\otimes\spl\otimes\spd\otimes\smd \to 
\spl\otimes\smd. \]
Here, we obtain another diagram that tells us how to interpret this 
mapping in global language.
\begin{prop}
\label{prop:mapsforwbar}
The following diagram commutes.
\begin{equation}
    \begin{CD}
        \spd\otimes\spl\otimes\spd\otimes\smd @>\eps^{+}>> 
        \spl\otimes\smd \\
        @VV\cong V            @VV\cong V \\
        ((X\times\RR)\oplus\wedge^{2}_{+}TX)\otimes TX@>C>> TX
    \end{CD}
    \label{eq:mapsforwbar}
\end{equation}
where the mapping \( C \) is given on a fiber by
\begin{equation}
    C((a, \omega)\otimes v) = av + i_{v^{*}}\omega.
\end{equation}
\end{prop}

\proof To prove \propref{mapsforw} we need to show that the bottom arrow 
in \eqref{mapsforw} is indeed given by 
the obvious projection.
We use \eqref{euclbasis} and \eqref{wedgetohomlower} to compute
\begin{align*}
    v_{1}\otimes v_{2} &= ( \eldl{1}{1} +  \eldl{2}{2})
                \otimes (i\eldl{1}{1} - i\eldl{2}{2}) \\
                &\stackrel{\eps^{+}}{\mapsto} 
                -i\ell{1}{2} - i\ell{2}{1} \\
                &= v_{1}\wedge v_{2} + v_{3}\wedge v_{4} \\
    v_{1}\otimes v_{3} &= ( \eldl{1}{1} +  \eldl{2}{2})
                \otimes (\eldl{1}{2} - \eldl{2}{1}) \\
                &\stackrel{\eps^{+}}{\mapsto} 
                \ell{1}{1} + \ell{2}{2} \\
                &= v_{1}\wedge v_{3} - v_{2}\wedge v_{4} \\
    v_{1}\otimes v_{4} &= ( \eldl{1}{1} +  \eldl{2}{2})
                \otimes (i\eldl{2}{1} + i\eldl{2}{1}) \\
                &\stackrel{\eps^{+}}{\mapsto} 
                i\ell{1}{1} - i\ell{2}{2} \\
                &= v_{1}\wedge v_{4} + v_{2}\wedge v_{3} \\
\end{align*}
with similar formulas for \( v_{2}\otimes v_{1} \) etc.  We also get
\begin{align*}
    v_{1}\otimes v_{1} &= ( \eldl{1}{1} +  \eldl{2}{2})
        \otimes (\eldl{1}{1} + \eldl{2}{2}) \\
        &\stackrel{\eps^{+}}{\mapsto} \ell{1}{2} - \ell{2}{1} \\
        &= 1 \in\RR
\end{align*}
with identical formulas for \( v_{i}\otimes v_{i} \), \( i=2, 3, 4. 
\) This completes the proof of \propref{mapsforw}. 
\propref{mapsforwbar} is proved with a similar computation.

The \( X\times\RR \) component of the space on the bottom of 
\eqref{mapsforw} is just the 
trivial bundle spanned by the identity section of \( T^{*}X\otimes TX 
\), followed by using the metric to map \( T^{*}X\to TX \).  So, what 
we have learned is that the unbarred spinors \(\lambda, \chi \), in twisted language, 
can be built as follows.  The image of the \odijcjcov{i}{c} under the 
vertical map in the diagram is simply the horizontal lift of \( D_{1} 
\), which we'll denote by \doc.  The image of the \dievencov{d}{a} is 
just the horizontal lift of de Rham \( d \), which 
we usually denote by \nab.  Let us decompose the projection on the 
bottom of the diagram with the maps
\begin{gather}
    \epsilon_{0}: TX\otimes TX\to X\times \RR \\
    \epsilon_{2}: TX\otimes TX\to \wedge^{2}_{+}TX.
\end{gather}
Then if we define
\begin{gather}
    W_{0} = \epsilon_{0}[\doc, \nab] \label{eq:w0}\\
    W_{2} = \epsilon_{2}[\doc, \nab] \label{eq:w2}
\end{gather}
\begin{gather}
    \psi_{0} = i^{*}W_{0} \label{eq:psi0}\\
    \psi_{2} = i^{*}W_{2} \label{eq:psi2}
\end{gather}
we have met two goals.  We have defined two component fields in 
global 
language on \( SX \), but we have also proved with \propref{mapsforw} 
that these two 
components can be rewritten in local coordinates as the \( \lambda \) 
and \( \chi \) we saw before.

We see from \eqref{mapsforwbar} that the restriction of \( C \) to the
subspace \( (X\times\RR)\otimes TX\cong TX \) has the same image as
all of \( C \).  So, we can build the barred spinors in a global way 
by forming the bracket \( [\dzc, \nab]. \)
\begin{gather}
    W_{1} = -\frac{1}{2}[\dzc, \nab]
    \label{eq:w1} \\
    \psi_{1} = i^{*}W_{1}
    \label{eq:psi1}.
\end{gather}

Next define
\begin{gather*}
    \Phi = -[\dzc,\dzc] \\
    \bar{\Phi} = -\frac{1}{4}\epsilon_0[\doc,\doc] \\
    \intertext{and}
    \phi = i^*\Phi \\
    \bar{\phi} = i^*\bar{\Phi}.
\end{gather*}
\begin{prop}
    \( \phi \) is the twisted version of \( \bar{\sigma} \) and \( 
    \bar{\phi} \) is the twisted version of \( \sigma \).
\end{prop}
\proof We compute that
\begin{align*}
    \Phi &= -[\dzc,\dzc]\\
    &=-[\mathcal{D}^{1}_{1}+\mathcal{D}^{2}_{2},
    \mathcal{D}^{1}_{1}+\mathcal{D}^{2}_{2}] \\
    &= -2[\mathcal{D}^{1}_{1},\mathcal{D}^{2}_{2}],
\end{align*}
which agrees with \eqref{sigma3}.  Similarly,
\begin{align*}
    -4\bar{\Phi} &= -\frac{1}{4}\epsilon_{0}[\doc,\doc]\\
    &=[\two{\cdibar{1}}-\one{\cdibar{2}},\two{\cdibar{1}}-
      \one{\cdibar{2}}] 
    +[i\two{\cdibar{1}}+i\one{\cdibar{2}},i\two{\cdibar{1}}+i\one{\cdibar{2}
     }]\\
    &+[\two{\cdibar{2}}+\one{\cdibar{1}},\two{\cdibar{2}}+\one{\cdibar{1}}]
    +[i\two{\cdibar{2}}-i\one{\cdibar{1}},i\two{\cdibar{2}}-i\one{\cdibar{1}}]\\
    &=4[\two{\cdibar{2}},\one{\cdibar{1}}]-4[\two{\cdibar{1}},\one{\cdibar{2}}],
\end{align*}
which is the sum of \eqref{sigma1} and \eqref{sigma2}, 
completing the proof.  However, if we compute
\begin{align*}
    -[\dtc,\dtc] &= [-i\mathcal{D}_{1}^{1}+i\mathcal{D}_{2}^{2},
                     -i\mathcal{D}_{1}^{1}+i\mathcal{D}_{2}^{2}]
                   +[\mathcal{D}_{2}^{1}-\mathcal{D}_{1}^{2}, 
                     \mathcal{D}_{2}^{1}-\mathcal{D}_{1}^{2}]\\
                  & +[-i\mathcal{D}_{2}^{1}-i\mathcal{D}_{1}^{2}, 
                     -i\mathcal{D}_{2}^{1}-i\mathcal{D}_{1}^{2}] \\
                 &= -2[\mathcal{D}^{1}_{1}, \mathcal{D}^{2}_{2}] + 
                   4[\mathcal{D}_{2}^{1},\mathcal{D}_{1}^{2}]
\end{align*}
then we see that it is also very natural to form the object
\[ -[\dzc,\dzc]-[\dtc,\dtc], \]
which agrees with the sum of \eqref{sigma3} and \eqref{sigma4}. 
However, we do not adopt this alternate definition of \( \Phi \).

We now proceed to discuss the fact that in flat space there are three
auxiliary fields.  The natural guess is to find some self-dual
two-vector field that can be written using the twist as these three
fields.  If we name this single twisted auxiliary field by the name \(
E_{2} \) then we claim
\begin{prop}
    \label{prop:auxiliaryequivalence}
    \( E_{2} = i^{*}\frac{1}{2}\epsilon_{2}(\doc[\dzc,\nab]) \) 
is the twisted version of the auxiliary fields \( E \) and \( F \).
\end{prop}
\proof We rewrite \( E_{2} \) as
\begin{align*}
    E_{2} &= i^{*}\frac{1}{2}\epsilon_{2}(\doc([\dzc,\nab])) \\
          &= -i^{*}\epsilon_{2}(\doc W_{1}) \\
            &= -i^{*}\epsilon_{2}(\doc\doc \Phi)
\end{align*}
and then consider what \( \epsilon_{2} \) does to \( \doc\doc. \)  If 
we 
label the four components of \doc\ in a coordinate chart by \( 
\doc^{i} 
\) then we can compute that
\begin{align*}
    \dd_{1}^{1}\dd_{1}^{2}+\dd_{1}^{3}\dd_{1}^{4} &= 
       (\dudlc{1}{1} + \dudlc{2}{2})(\dudlc{1}{2} - \dudlc{2}{1})
      -(i\dudlc{1}{1} - i\dudlc{2}{2})(i\dudlc{1}{2} + i\dudlc{2}{1}) \\
      &= \dudlc{1}{1}\dudlc{1}{2} - \dudlc{2}{2}\dudlc{2}{1}
\end{align*}
and so checking with \eqref{defofcomponentsinr482} we see that one of 
the three components of \( \epsilon_{2}\doc\doc\Phi \) is \( 
\bar{F}-F. 
\)  Similar computations reveal that the second component, \( 
D_{1}^{1}D_{1}^{3}-D_{1}^{2}D_{1}^{4}, \) is \( 
F+\bar{F} \) and the third, \( D_{1}^{1}D_{1}^{4}+D_{1}^{2}D_{1}^{3}, 
\) gives \( E \), which completes the proof.

In summary we have proved the following key result.
\begin{theorem}
    Let \pp\ be a principal \sut\ bundle over the supermanifold \( SX 
    \).  Let \( P\to X \) be the restriction of \pp\ to \( X \). 
    The space of semi-constrained superconnections on \pp\ is 
    isomorphic to the space of fields \( A, \phi, \bar{\phi}, 
    \psi_{0}, \psi_{1}, \psi_{2}, E_{2}, \) where \( A \) is an 
    ordinary connection on the restriction \( P\to X \) of \( \pp\to 
SX 
    \),  \( \phi \) and \( \bar{\phi} \)  are sections of \( \adp 
    \), \( \psi_{0} \in \Gamma(\Pi(X\times\RR)\otimes\adp)) \), \( 
    \psi_{1} \in \Gamma(\Pi TX\otimes\adp) \), \( 
    \psi_{2} \in \Gamma(\Pi\wedge^{2}_{+}TX\otimes\adp) 
    \), and \( E_{2}\in \Gamma(\wedge^{2}_{+}TX\otimes\adp). \)
\end{theorem}
\proof It suffices to work in a coordinate patch, where by the
preceding discussion the theorem reduces to its \( N=2 \) flat space
counterpart \thmref{semiconstrained}.

\subsection{The action of \mytexorpdfstring{$Q_{0}$}{Q0}}

\label{sec:actionofq0}
We will now go about computing the vector field on \sa\ that is
induced by \( Q_0 \).  What we mean is that the vector field \( Q_0 \)
on \( SX \) acts on functions and bundle sections by covariant
differentiation, and so it acts on the points of the space \sa.  The
infinitesimal form of this action is again a vector field and we are
going to try to express it in terms of components.  Let \( \eta \) be
an odd parameter.  If \aa\ is a semi-constrained superconnection then
denote by \( \xi \) the diffeomorphism generated by the even vector
field \( \eta\qzc \).  To get formulas in components, then, we are
searching for the components of \( \xi\aa.\) These components are
defined using the \( \mathcal{D}_i \), which all commute with \(\xi\). 
Moreover, when restricted to the even submanifold \(P\subset \pp\) the
vector fields \dzc\ and \qzc\ agree with each other.  This all implies
that we can compute the action of \qzc\ by using \dzc\ instead.  And
so, our approach will simply be to hit the component fields with \dzc\
on the left and rewrite them in terms of each other after some dust
settles.  A key tool will be the super Bianchi identity, which
is the Bianchi identity with parity taken into account.
\begin{theorem}[Bianchi]
Let \ff\ be the curvature of a connection \aa\ on a principal bundle
\( \pp\to SX \).  Let \( X, Y, Z \) be vector fields on \( SX \).  Let
\( \hat{X}, \hat{Y}, \hat{Z} \) denote the horizontal lifts to \pp.
If \( \pi(X) \) denotes the parity of the vector field \( X \) then 
\begin{align*}
    0=(\hat{X}{\ff})(Y, Z) &+ 
    (-1)^{\pi(X)\pi(Y)+\pi(X)\pi(Z)}(\hat{Y}{\ff})(Z, X) \\&+
    (-1)^{\pi(Z)\pi(X) + \pi(Z)\pi(Y)}(\hat{Z}{\ff})(X, Y),
\end{align*}
where the covariant derivative of a two-form is given by the super 
formula
\[ (\hat{X}{\ff})(Y, Z) = 
\hat{X}(\ff(Y, Z)) - \ff([X,Y], Z) + (-1)^{\pi(X)\pi(Y)} \ff(Y, 
[X,Z]).\]
\end{theorem}

We apply this theorem as follows.  The identity on the three vector 
fields \( D_{0} \), \( D_{0} \) and \( d \) yields
\[
\bianchioddoddeven{D_{0}}{D_{0}}{d}{\dzc}{\dzc}{\nab}.
\]
In this expression, the second and sixth terms vanish due to \(
D_{0}^{2} = 0 \) and the third, fifth, eighth and ninth vanish due to
the fact that the connection is semi-constrained.  This leaves us with
\[ \dzc(\ff(D_{0}, d)) - \dzc(\ff(d, D_{0})) + \nab(\ff(D_{0}, 
D_{0})) = 0,\]
which gives on restriction to \( X \)
\begin{equation}
    \boxed{\dzc \psi_{1} = -\nab \phi.}
    \label{eq:d0psi1}
\end{equation}

Next we examine the identity for three copies of \( D_{0} \).
\[
\bianchioddoddodd{D_{0}}{D_{0}}{D_{0}}{\dzc}{\dzc}{\dzc}
\]
which immediately becomes 
\begin{equation}
    \boxed{\dzc \phi = 0.}
    \label{eq:d0phi}
\end{equation}

Next we work with \( D_{0} \), \( D_{1} \) and \( d \) to obtain
\[
\bianchioddoddeven{D_{0}}{D_{1}}{d}{\dzc}{\doc}{\nab}
\]
whose third, fifth, seventh, eighth and ninth terms vanish because 
the 
connection is semi-constrained.  Using the fact that
\[ [D_{1},D_{0}] = [D_{0},D_{1}] = d \]
gives
\[ \dzc((\ff(D_{1}, d)) - \doc(\ff(d, D_{0})) = 0. \]
This yields the equation
\begin{align}
    \dzc \psi_{2} &= i^{*}
    \epsilon_{2}(\doc(\ff(d, D_{0}))) \nonumber\\
    &= E_{2}.
    \label{eq:q0psi2}
\end{align}
so we have
\begin{equation}\boxed{\dzc\psi_2=E_2.}\end{equation}

Next we compute with \( D_{0} \), \( D_{1} \) and \( D_{1} \).
\[
\bianchioddoddodd{D_{0}}{D_{1}}{D_{1}}{\dzc}{\doc}{\doc}.
\]
the fourth and seventh terms vanish by the semi-constrained 
condition.  We are going to use this equation to compute \( 
\dzc\bar{\phi} 
\) and so we need to take \( \epsilon_{0} \) of both sides.  \( 
\epsilon_{0} \) 
takes the trace of this bracket, and because individual 
components of \doc\ square to zero we kill the terms with \( 
[\doc,\doc] \).  This leaves us with
\begin{equation}
    \boxed{\dzc\bar{\phi} = \psi_{0}.}
    \label{eq:d0phibar}
\end{equation}

Next let's work out \( \dzc A. \)  To compute a component 
of this one-form we'd examine the restriction to \( X \) of
\[ i(\nab_{x^{\mu}}) L(\dzc)A. \]
To get the global version of this, we simply replace the partial with 
\nab.
    \begin{align}
        i(\nab)L(\dzc)A &= L(-\dzc)i(\nab)A - i(2[\dzc,\nab])A \\
        &= 0 - 2[\dzc,\nab] \\
        &= W_{1}
    \end{align}
so that we obtain
\begin{equation}
    \boxed{\dzc A = \psi_{1}.}
    \label{eq:d0a}
\end{equation}

This leaves us with the computations for \( \psi_{0} \) and \(
E_{2} \).  These are in the image of \dzc\ though, and we
can argue as follows.  If \( f \) is some component field, we can 
compute
\begin{align}
    i^{*}\dzc\dzc f &= i^{*}[\dzc,\dzc]f \\
                    &= i^{*}\nab_{F(D_{0},D_{0})}f \\
                    &= \nab_{-\phi}f \\
                    &= -[\phi,f].
\end{align}
Using this we obtain
\begin{gather}
    \boxed{\dzc \psi_{0} = -[\phi,\bar{\phi}]} \\
    \boxed{\dzc (F_{A}^{+}-E_{2}) = -[\phi,\psi_{2}].}
\end{gather}

The total result is then
\begin{equation}
    \label{eq:q0onfields}
    \boxed{
    \qzc\left(
    \begin{array}{c}
        A \\
        \psi_{1} \\
        \phi \\
        \bar{\phi} \\
        \psi_{0} \\
        \psi_{2} \\
        E_{2}
    \end{array}
    \right)
     = 
     \left(
     \begin{array}{c}
         \psi_{1} \\
         -\nab\phi \\
         0 \\
         \psi_{0} \\
         -\left[\phi,\bar{\phi}\right] \\
         E_{2} \\
         -\left[\phi, \psi_{2}\right]
     \end{array}
     \right)_{(A, \psi_{1}, \phi, \bar{\phi}, \psi_{0}, \psi_{2}, 
E_{2})}
    }
\end{equation}

This is a vector field on \sa, and so can be used as a differential 
operator on functions on \sa.  We have 
decomposed the infinite dimensional space \sa\ into seven subspaces, 
\begin{equation}
    \label{eq:structureofsa}
    \sa = \aa \times\Pi \Omega^{1} \times \Omega^{0} \times \Omega^{0}
    \times \Pi \Omega^{0} \times \Pi \Omega^{2}_{+} \times
    \Omega^{2}_{+}
\end{equation}
(where we omit the \( (X; \adp) \)'s from the notation for clarity). 
Suppose \( f \) is a function on \sa, and we would like to compute the
derivative of \( f \) using \( \qzc \), i.e. \( \qzc f 
\). 
Suppose further that we have an explicit expression for \( f \) as a
combination of various component fields.  In order to compute a
similarly explicit expression for \( \qzc f \), we use the Leibnitz
rule and the chain rule, and then ask the question ``what is \( \qzc 
\)
evaluated on an individual component field?''  One computes this
derivative by taking the corresponding component of \( \qzc \).  

An analogous situation
is the following.  Suppose we work on a finite-dimensional manifold \(
X \) and use local coordinates \( x^{\mu}, \mu=1, \ldots, n \) to
express a computation.  Let \( f(x^{1}, \ldots, x^{n}) = x^{i} \) for
some fixed \( i \) in these coordinates.  If \( V =
\sum_{k}a^{k}(x^{1},\ldots,x^{n})\frac{\partial}{\partial x^{k}} \) is
a vector field in this patch, then \( Vf(x^{1}, \ldots, x^{n})   = 
\) \(
a^{i}(x^{1}, \ldots, x^{n}),\)  the \( i \)th component of \( V \).

\subsection{The \action\ after the twist}
\label{sec:twistedaction}
We will twist the formula \eqref{e4action}.  Many of the terms 
there do not respond to the twist, which affects only the fermions 
and auxiliary fields as we have seen.  However, note that in our 
notation \( \sigma \) is \( \bar{\phi} \) and \( \bar{\sigma} \) is \( 
\phi \).  With just this we find that the twisted action has the terms
\[ \frac{1}{g^{2}}\left\{-|F_{A}|^{2} + 
-i\ip{\bar{\phi}}{d_{A}^{*}d_{A}\phi} -i 
|E_{2}|^{2}\right\} + \frac{\theta}{16\pi^{2}}\langle F_{A},F_{A}\rangle . \]

It remains to twist the fermionic terms.  First we examine
\[ \langle{\lambda}\odiraca_{A}\bar{\lambda}\rangle
    + \langle{\chi}\odiraca_{A}\bar{\chi}\rangle. \]
\begin{lemma}
    In local coordinates on \( SX \) we have
    \begin{equation}
        \label{diractoforms}
        \langle{\lambda}\odiraca_{A}\bar{\lambda}\rangle
        + \langle{\chi}\odiraca_{A}\bar{\chi}\rangle
        = \ip{\psi_{0}}{d_{A}^{*}\psi_{1}} 
        + \ip{\psi_{2}}{d_{A}^{+}\psi_{1}}
    \end{equation}
\end{lemma}
\proof We express the right hand side in spinorial coordinates to
prove the lemma.  By way of motivation, examine \eqref{diracpairing}. 
The element being hit with the two epsilon tensors is an element of \[
\spd\otimes\spl\otimes \spd\otimes\smd\otimes\smd\otimes \spl \] and
the \( \epsilon \)'s are contracting the two \spd\ spaces and the two
\smd\ spaces.  We formed a picture of these two contractions in
Propositions~\ref{prop:mapsforw} and \ref{prop:mapsforwbar}.  For
instance, the \( \epsilon^{-} \) contraction on the \smd\ spaces will
combine the \( \nabla_{a\dot{b}}\lambda_{\dot{b}} \) (and its twin \(
\nabla_{a\dot{b}}\chi_{\dot{b}} \) which is not separate in this
context) by mapping to \( (X\times\RR)\oplus\wedge^{2}_{+}TX. \)
Clearly this will produce \( d_{A}^{*}\psi_{1} \) and \(
d_{A}^{+}\psi_{1} \).  Recalling \eqref{euclbasis}, \eqref{vtohom} and
\eqref{wedgetohom} we compute
\[
\begin{split}
2d_{A}^{+}\psi_{1} = &\left(
\begin{array}{c}
    (\dieven{1}{1}+\dieven{2}{2})(-i\cbi{1}-i\lbi{2})
    -(i\dieven{1}{1}-i\dieven{2}{2})(-\cbi{1}+\lbi{2})\\
%
    (\dieven{1}{1}+\dieven{2}{2})(-\cbi{2}-\lbi{1})
    -(\dieven{1}{2}-\dieven{2}{1})(-\cbi{1}+\lbi{2})\\
%
    (\dieven{1}{1}+\dieven{2}{2})(-i\cbi{2}+i\lbi{1})
    -(i\dieven{1}{2}+i\dieven{2}{1})(-\cbi{1}+\lbi{2})
\end{array}\right. \\
& \qquad\left.\begin{array}{c}
    +(\dieven{1}{2}-\dieven{2}{1})(-i\cbi{2}+i\lbi{1})
    -(i\dieven{1}{2}+i\dieven{2}{1})(-\cbi{2}-\lbi{1}) \\
%
    -(i\dieven{1}{1}-i\dieven{2}{2})(-i\cbi{2}+i\lbi{1})
    +(i\dieven{1}{2}+i\dieven{2}{1})(-i\cbi{1}-i\lbi{2}) \\
%
    +(i\dieven{1}{1}-i\dieven{2}{2})(-\cbi{2}-\lbi{1})
    -(\dieven{1}{2}-\dieven{2}{1})(-i\cbi{1}-i\lbi{2})
\end{array}\right)
\end{split}\]
To express \( \psi_{2} \) in spinor coordinates, we use 
\eqref{wedgetohom} together with the fact that elements with upper 
index 1 are called \( \chi \) and with upper index 2 are called \( \lambda 
\) (see \eqref{lambda} and \eqref{chi}).  We get
\[ 4\psi_{2} = \left(
\begin{array}{c}
    -i\chi_{1}+i\lambda_{2} \\
    \chi_{2}-\lambda_{1} \\
    -i\chi_{2}-i\lambda_{1}
\end{array}\right).\]
Similarly,
\[ 4\psi_{0} = \chi_{1}+\lambda_{2} \]
and 
\[ \begin{split}
2d_{A}^{*}\psi_{1} = & (\dievencov{1}{1}+\dievencov{2}{2})
                      (-\cbi{1} + \lbi{2})+
                      (i\dievencov{1}{1}-i\dievencov{2}{2})
                      (-i\cbi{1} - i\lbi{2}) \\
                    & + (\dievencov{1}{2}-\dievencov{2}{1})
                      (-\cbi{2} - \lbi{1}) +
                      (i\dievencov{1}{2}+i\dievencov{2}{1})
                      (-i\cbi{2} + i\lbi{1})
\end{split}                  
\]
Computing \( \ip{\psi_{0}}{d_{A}^{*}\psi_{1}} +
\ip{\psi_{2}}{d_{A}^{+}\psi_{1}} \) is now a matter of combining 
these expressions and cancelling half of the terms, leaving us with 
the desired quantity.

Next we work with the terms involving brackets of spinors.  Something 
surprising will result --- a term that will not play a role in the 
geometrical picture that emerges in the next section.
\begin{lemma}
    \label{lemma:brackets}
    In local coordinates on \( SX \) we have
    \begin{align}
        \epsilon^{ab}[\lambda_{a},\chi_{b}] &= \frac{1}{4}[\psi_{2},\psi_{2}] + 
        \frac{1}{4}[\psi_{0},\psi_{0}] \\
        \epsilon^{\dot{a}\dot{b}}[\lbi{a},\cbi{b}] &= 
        \epsilon_{0}[\psi_{1},\psi_{1}].
    \end{align}
\end{lemma}
\proof We compute
\begin{align*}
    \epsilon_{0}[\psi_{1},\psi_{1}] =& 
    -\frac{1}{4}([-\cbi{1}+\lbi{2}, -\cbi{1}+\lbi{2}] +
    [-i\cbi{1}-i\lbi{2}, -i\cbi{1}-i\lbi{2}] \\ &+
    [-\cbi{2}-\lbi{1}, -\cbi{2}-\lbi{1}]+
    [-i\cbi{2}+i\lbi{1}, -i\cbi{2}+i\lbi{1}]) \\
    =& [\lbi{1},\cbi{2}] - [\lbi{2},\cbi{1}].
\end{align*}
And using the fact that the bracket uses the structure of the wedge 
product on forms, and that the components of \( \psi_{2} \) wedge to 
zero except against themselves, we obtain
\begin{align*}
    \frac{1}{4}[\psi_{2},\psi_{2}]+\frac{1}{4}[\psi_{0},\psi_{0}] =& 
    \frac{1}{4}([-i\lambda_{1} + i\chi_{2}, -i\lambda_{1} + i\chi_{2}] +
    [\lambda_{2} - \chi_{1}, \lambda_{2} - \chi_{1}] \\
    &+ 
    [-i\lambda_{2} - i\chi_{1}, -i\lambda_{2} - i\chi_{1}] +
    [\lambda_{1} + \chi_{2}, \lambda_{1} + \chi_{2}]) \\
    =& [\lambda_{1},\chi_{2}]-[\lambda_{2},\chi_{1}].
\end{align*}
This completes the proof and thus we have computed
\begin{align}
    \int_{\sa} 
  \exp\Big(\frac{1}{g^{2}}
    \Big( 
    &-\frac{1}{2}|F_{A}|^{2}  
    - i\ip{\bar{\phi}}{d_{A}^{*}d_{A}\phi}
    +i\ip{\psi_{0}}{d_{A}^{*}\psi_{1}} 
    + i\ip{\psi_{2}}{(d_{A}\psi_{1})^{+}}   \nonumber\\
    &+ i\ip{\psi_{2}}{[\phi, \psi_{2}]}
    + i\ip{\phi}{[\psi_{0},\psi_{0}]}
    + i\ip{\bar{\phi}}{[\psi_{1}, *\psi_{1}]} \\
    &- i|E_{2}|^{2}\Big)
    + \frac{\theta}{16\pi^{2}}\langle F_{A}\wedge F_{A}\rangle
    \Big)
\end{align}
If we tweak the parameter \( \theta \), we can obtain the sum
\[ -\frac{1}{2g^{2}}|F_{A}|^{2} - \frac{1}{2g^{2}}\langle F_{A}\wedge F_{A}\rangle \]
which becomes
\[ -\frac{1}{g^{2}}|F_{A}^{+}|^{2} \]
This particular value for \( \theta \) will be fixed from now on, for
it facilitates the geometric interpretation we will dwell on
presently.  Note that with this alteration the whole action has an
overall coefficient of \( \frac{1}{g^{2}}.  \) This is the coupling
parameter for this physical theory, and when written outside the action
it acts like Planck's constant \( h \).  Namely, we can see directly
(if path integration makes sense) that if the coupling becomes
vanishingly small then the minima of \( S \) are heavily weighed in a
path integral computation, and we approach a classical limit.  We will
prefer a different interpretation for the coupling parameter and so we
scale some of the fields as follows
\begin{align*}
    \phi &\mapsto g^{2}\phi \\
    \psi_{0} &\mapsto   g^{2}\psi_{0}\\
    \psi_{2} &\mapsto   g^{2}\psi_{2} \\
    E_{2}    &\mapsto   gE_{2},
\end{align*}
producing the formula we will use going forward:
\begin{align}
    \label{eq:twistedaction}
        \int_{\sa} 
      \exp \Big( 
        &-\frac{1}{g^{2}}|F_{A}^{+}|^{2}  
        - i\ip{\bar{\phi}}{d_{A}^{*}d_{A}\phi}
        +i\ip{\psi_{0}}{d_{A}^{*}\psi_{1}} 
        + i\ip{\psi_{2}}{(d_{A}\psi_{1})^{+}} \\
        &+ g^{2}i\ip{\psi_{2}}{[\phi, \psi_{2}]}
        + g^{2}i\ip{\phi}{[\psi_{0},\psi_{0}]}
        + i\ip{\bar{\phi}}{[\psi_{1}, *\psi_{1}]}
        - i|E_{2}|^{2}\Big)
\end{align}

We make one final remark about this computation.  It should in 
principle be possible to compute the twisted action directly on \( SX 
\), perhaps using a multiple of
\[ \int_{\Pi TX}\mathrm{Tr}\, \Phi^{2} - 
\int_{\Pi((X\times\RR)\oplus\wedge_{2}^{+}TX)}\mathrm{Tr}\,\bar{\Phi}^{2}. 
\]
which corresponds to \eqref{superaction}.  To compute this in 
components, one would hit each integrand with an appropriate 
differential operator.  For example, the first odd integral could be 
carried out by hitting \( \mathrm{Tr}\,\Phi^{2} \) with \( (D_{1})^{4} 
\), interpreted in an appropriate sense.  Similarly, the second 
integral could be carried out with the help of \( D_{0}\circ (D_{2})^{3} 
\) where the cube is perhaps interpreted to mean the determinant on the 
third tensor power of the 3-dimensional bundle \( \wedge_{2}^{+}TX. 
\)  This computation should be straightforward once the meaning of 
these operators is sorted out.  Some insight into \( SX \) is sure to 
be gained by this exercise.

\section{The polynomial invariants}

The definitions of the component fields of semi-constrained
superconnections give a decomposition of \( \sa \).  A central result
of this paper is that this decomposition can be viewed as a rich
algebraic structure living on the usual space of connections.  Without
having ever mentioned the ASD equations or the action of the group of
gauge transformations, we will find that in a formal sense these are
\emph{automatically called for} by the structure of \( \sa \).  

Let \( P \) be a principal \( G \)-bundle over a base \( X \) and let
\( V \) be a \( 2n \)-dimensional representation of \( G \).  Form the
associated vector bundle \( E = P\times_{G}V \).  On this vector
bundle there is a Thom class \( \thom\in H^{2n}_{c}(E) \) in compactly
supported cohomology.  It has maximal degree along the fibers, and so
is fully ``vertical.''  One can pull back a representative of \thom\
to \( X \) by the zero section \( s \) and obtain a representative \(
e \) of the Euler class of \( E \).  If one pulls \thom\ back by a
nonzero section \( s \), one can interpret the pullback \(
s^{*}(\thom) \) as the Poincar\'e dual to the zero set \( Z_{s} \) of
\( s \).  And so, to integrate a differential form \( \omega \) over
\( Z_{s} \) one can integrate \( \omega\wedge s^{*}(\thom) \) over all
of \( X \).

Mathai and Quillen~\cite{mathaiquillen} introduced a representative \(
\thom_{A} \) for the Thom class that lives in the \( G \)-equivariant
cohomology of \( P\times V. \) The \( A \) denotes a connection on \(
P \), which is used in the construction.  In fact, they write an
element of the Cartan algebra of \( V \), which is an algebraic model
of equivariant cohomology, and then use the connection to map it to an
equivariant differential form on \( P\times V \), using the Weil
homomorphism.  Mathai and Quillen showed that if \( s \) is an
arbitrary section of \( E \), then \( s^{*}\thom_{A} = e_{s, A} \) is
a representative for the Euler class, and is independent of both \( A
\) and \( s \).  To be totally explicit, they write
\begin{equation}
    \label{eq:mqform}
    e_{s,A} = \frac{1}{(2\pi)^{n}}\int d\psi\,e^{-\|s\|^{2} + 
    \frac{1}{2}\ip{\psi}{\phi\psi} + i\ip{ds}{\psi}}
\end{equation}
where \( \psi \) is an element of \( \Pi V \).  The object \( \phi \) 
is an element of the equivariant cohomology, and under the Weil 
homomorphism it maps to \( F_{A} \), which we will discuss a little 
later.  Note that this element has \emph{rapid decrease} along the 
fiber \( V \), but is not compactly supported.  In fact, the 
inclusion of compactly supported forms into forms with rapid 
decrease induces an isomorphism of cohomology.

Taking \( s=0 \) produces \( \mathrm{Pfaff}(F_{A}) \), which restates
the Gauss-Bonnet theorem.  Taking \( s \) nonzero and multiplying by a
constant \( \gamma \) to get \( \gamma s \) and then taking \(
\gamma\to\infty \) \emph{localizes} \( e_{s,A} \) to the zero set of
\( s \).  This can be proven by approximating \eqref{mqform} with the
method of steepest descent.  

Also relevant for us is a modification of this picture that lets us 
work upstairs.  Let \( s \) be a section of \( E \), and suppose we 
want to compute
\[ \int_{Z_{s}}\omega \]
for some form \( \omega \).  We know we can instead work with
\[ \int_{X}\omega\wedge e_{s, A}. \]  
However, we can further enlarge the space we integrate over to \( E \) 
if we can find an appropriate differential form that has maximal degree 
along the 
fibers of \( P \) and that integrates to 1 on a fiber.  Such a form is 
called a 
\emph{projection form}, and if we call it \( \eta_{\mathrm{proj}} \) 
then we have
\[ \int_{Z_{s}}\omega = \int_{E}\omega\wedge e_{s,A}\wedge 
\eta_{\mathrm{proj}}. \]

It is familiar in Donaldson theory that the
ASD moduli space can be defined as the zero set of
the section \( F_{A}^{+}:\aa/\gg\to \Omega^{2}_{+}(X; \adp).  \) If
there were such an object as a Thom class in this infinite-dimensional
context, we could hope that the pullback by \( F_{A}^{+} \) would be
in some sense Poincar\'e dual to the ASD moduli space.  Surely such a
geometrical construction could be carried out mathematically, but it has
not yet been done.  The problem is that the space \bb\ of connections 
modulo gauge transformations is infinite-dimensional and the fiber of 
the vector bundle, \( \Omega^{2}_{+}(X; \adp) \) is also 
infinite-dimensional.  In addition, the group \gg\ has infinite 
dimension, so the concept of the projection form as a 
``top-dimensional'' form along the fibers of \( \aa\to\bb \) does not 
make sense.  Nonetheless, if we ignore these issues we will see that a 
straightforward application of the above construction to \( F_{A}^{+} \) 
produces the twisted action \eqref{twistedaction}.

So the simple twisting operation has brought us from a physical
supersymmetric theory all the way to the ASD moduli space, equipped
with an Euler class to help us do intersection theory.  All that is
missing is Donaldson's \( \mu \)-map, which has a beautiful
manifestation in this context, as we will see below.

Much of this treatment of the Mathai-Quillen form and the projection
form is based on \cite{cordes}.  The original insight into the
geometry underlying the action is in Atiyah and Jeffrey's paper
\cite{atiyahjeffrey}.  The following account differs from Atiyah and
Jeffrey's, however, in two important respects.  First of all, we build
the geometrical constructions from the structure of \sa\ itself, using
the operator \qzc\ and the component fields to prove that the
equivariant cohomological data we need is encoded in the twisted
supersymmetry.  This is a very important observation, as it uses the
twist to show that the Mathai-Quillen and projection forms naturally
arise from supersymmetry, and so \emph{motivate} doing Donaldson
theory rather than just \emph{imitating} Donaldson theory.  Thinking
of Donaldson theory as an outgrowth of twisted supersymmetry may
eventually prove to be useful for gaining additional insights
about smooth structures on 4-manifolds.  The second departure from 
Atiyah and Jeffrey's work is that we will try to de-emphasize the 
interpretation of the path integral as a representation of a 
nonexistent Euler class.  Instead, we will discuss the physical 
approach to path integration and show how the localization to the ASD 
moduli space is obtained by examining the classical limit of the 
quantum theory.  Strengthening the link with physics fits into our 
overarching strategy of initiating an investigation into Witten's 
Conjecture, but the reader should be clear on one point: understanding 
what an Euler class is in infinite dimensions will shed light on both 
the Donaldson invariants and on path integrals, and so we are not 
advocating that mathematicians should neglect to sort those ideas out.

\subsection{The algebraic structure of \mytexorpdfstring{$\sa$}{SA}}

We define two subspaces of \sa.
\begin{align}
    L(\aa) &= \Omega^{0} \times \Omega^{2}_{+} \times \Pi
    \Omega^{2}_{+} \\
    P(\aa) &= \aa \times\Pi \Omega^{1} \times \Omega^{0} \times \Omega^{0}
    \times \Pi \Omega^{0}
\end{align}
where the shared copy of \(\Omega^0\) is the one given by elements 
we have been calling \(\phi\).
(Sometimes we will want to use dual spaces of a few of these pieces
but we will feel free to switch to the dual spaces as needed.) 
Keeping this structure in mind, we will digress temporarily to treat
more carefully the two finite-dimensional geometrical ideas, the
Mathai-Quillen form, and the projection form.  Our presentation of these two 
forms relies on the algebraic structure of the Cartan model for 
equivariant cohomology.  Then we will return to
\sa\ and see that we have the same algebraic picture present, in the 
guise of the vector
field \qzc\ and in the twisted action \eqref{twistedaction}.

\subsubsection{The Mathai-Quillen form}

In \cite{mathaiquillen}, Mathai and Quillen constructed a
representative for the Euler class of a vector bundle that is built 
from a
connection and a section.  The proved that their form was closed and
that its cohomology class depends neither on the section nor the
connection.  We will describe their construction now.  Let \( G \) be
a Lie group, let \( A \) be a principal \( G \)-bundle over a space \(
B \), and let \( E = A\times_{\rho}V \) be an associated \( n
\)-dimensional vector bundle, where \( \rho:G\to GL(V) \) is a given
representation.  We are using finite dimensional \( A, B\) and \( G
\), but their names should suggest that we will eventually apply these
ideas to the infinite dimensional spaces \aa, \bb\ and \gg.  

Let \( d\rho:\g\to \mathrm{Hom}(V)\cong \mathrm{Vect}(V) \) be the 
Lie algebra map to vector fields on \( V \). We will denote by \( 
i_{\phi} \) the contraction operator in the direction of \( 
d\rho(\phi) \). The Cartan algebra is the space \( 
S^{*}(\g^{*})\otimes 
\formsij{*}{V}, \) equipped with a differential \( d_{C} \) given by
\begin{align}
    d_{C}(\phi\otimes 1) &= 0 \\
    d_{C}(\phi\otimes\omega) &= 1\otimes (d-i_{\phi})\omega
\end{align}
where \( \phi \) is a generator of \( S^{*}(\g^{*}) \), and one 
extends this formula to the full algebra by the Leibnitz rule.  The 
cohomology of this complex computes the equivariant cohomology of \( V 
\), \( H^{*}_{G}(V). \)  In case \( G \) acts freely, one has \( 
H^{*}_{G}(V)\cong H^{*}(V/G), \) so this algebraic model is 
designed to help handle the cases where the action is not free.

To make closer contact with our work on superconnections, we can 
describe the Cartan algebra using a vector field on an odd space.
\begin{prop}
    Let \( V \) be a vector space with inner product, together with an
    action of a Lie group \( G \) (not necessarily linear).  Let a
    metric on \g\ be given.  Let \( \{v_{i}\} \) be a basis of \( V \)
    and let \( \{v^{i}\} \) be the dual basis.  Define \( CV 
     = \g\times\Pi TV \). Let a vector field \(
    Q \) on \( CV\) be given by 
    \[ Q\left(
    \begin{array}{c}
	\phi_{i} \\
	v_{j} \\
	\lambda_{k}
    \end{array}
    \right) = \left(
    \begin{array}{c}
	0 \\
	\lambda_{j} \\
	-L(\phi_{i})v_{k}
    \end{array}
    \right)_{(\phi_{i}, w_{j}, \lambda_{k})}
    \]
    where \( \phi_{i} \) is an element of a basis for \g\ and \(
    \lambda_{j} \) is the basis element of \( \Pi V\cong \Pi 
T_{v_{j}}V
    \) corresponding to \( v_{j} \).  Then on the space 
\(S^*(\g)\otimes 
    \formsij{*}{V}\), \( Q \) induces the 
    action of the Cartan differential.
    \label{prop:qinducescartan}
\end{prop}
\proof We have already discussed in \lemmaref{functionsforms} how \(
C^{\infty}(\Pi TV)\cong \formsij{*}{V} \) and so taking the
space of polynomial functions on \g, we have \(
S^{*}(\g^{*})\otimes \formsij{*}{V}\subset C^{\infty}(\g\times\Pi
TV).  \)  \( Q \) induces an action on this space by differentiation, 
and so to complete the proof we compute this induced action.  Let us 
denote the superspace analogue of a differential form \( \omega \) by 
\( 
\hat{\omega} \).  So if
\[ \omega = \phi^{\alpha}\cdot
\sum_{i_{1}<\cdots<i_{k}}a_{i_{1}\cdots i_{k}}(v) 
dv^{i_{1}}\wedge\cdots\wedge dv^{i_{n}} \]
is an element of the Cartan algebra, then the corresponding function 
on \( \g\times\Pi TV \) is
\[ \hat{\omega} = \phi^{\alpha}\cdot
\sum_{i_{1}<\cdots<i_{k}}a_{i_{1}\cdots i_{k}}(v) 
\lambda_{i_{1}}\cdots \lambda_{i_{n}}. \]

Let us compute the action of \( Q \) on \( \hat{\omega}. \)  We have
\begin{align}
    Q\hat{\omega} 
    &= \sum a_{i_{1}\cdots 
    i_{k}}(-1)^{\gamma+1}
    \lambda_{i_{1}}\cdots\lambda_{i_{\gamma-1}}
    (-L(\phi_{\alpha})v^{\gamma})
    \lambda_{i_{\gamma+1}}\cdots 
    \lambda_{i_{k}} \nonumber\\
    &+ \phi^{\alpha}\cdot\sum 
    \frac{\partial a_{i_{1}\cdots i_{k}}}{\partial v_{\gamma}}
    \lambda_{\gamma}\lambda_{i_{1}}\cdots 
\lambda_{i_{n}}\label{eq:qonfunctions}.
\end{align}
Note that we computed only for a generator of \( S^{*}(\g^{*}) \) but 
this suffices as both \( Q \) and the Cartan differential are 
extended 
in the same way (the Leibnitz rule) to more general elements. 
Under the correspondence with differential forms, \( \lambda_{i}\to 
dv^{i} 
\).  Also, by the Cartan formula, \( L = d\circ i + i\circ d \) and 
so 
\[ -L(\phi_{\alpha})v^{\gamma} = -i(\phi_{\alpha})dv^{\gamma}. \]
With these replacements, we can easily see from \eqref{qonfunctions} 
that \( Q\hat{\omega} \to 
(d-i(\phi_{\alpha}))\omega \), completing the proof.

We deliberately avoided using linearity of the \( G \)-action above, 
in order to be a little more general.  However, in the case of a 
linear action (a representation) the action of \( \phi \) 
on an element of \( V \) is just the vector field 
\( \phi(v)_{v} \).  

Let us examine the space \( L'(A) = \g\times\Pi TV\times \Pi
V^{*}\times V^{*}.  \) (The number of components differs from the
definition of \( L(\aa) \) above, hence the primed notation; the \(
\Pi TV \) part of \( L'(A) \) should be considered ``extra'' and we
will see at the end of the section why its presence is not needed to
discuss the Donaldson invariants.)  We will install on this space the
vector field
\begin{equation}
    Q\left(
    \begin{array}{c}
	\phi \\
	v_{j} \\
	\lambda_{k} \\
	\psi^{l} \\
	E^{m}
    \end{array}
    \right) = \left(
    \begin{array}{c}
	0 \\
	\lambda_{j} \\
	-\phi(v_{k}) \\
	E^{l} \\
	-\phi(\psi^{m})
    \end{array}\right)_{(\phi_{i},v_{j},\lambda_{k},\psi^{l},E^{m})}.
    \label{eq:qlocalization}
\end{equation}
We will construct a special function \( \Psi_{L} \) on this space as 
follows.  We will then integrate the exponential of the function \( 
Q\Psi_{L} \) over the \( E \) and \( \psi \) variables, and we 
shall 
point out that the remaining function is Mathai and Quillen's element 
of the Cartan algebra.  Let us proceed.  We set
\begin{equation}
    \Psi_{L}(\phi, v, \lambda, \psi, E) = -i\psi(v) 
    -\ip{\psi}{E}_{V^{*}}
    \label{eq:psilocalization}
\end{equation}
and then obtain
\begin{equation}
    \Phi_{L} = Q\Psi_{L} = -iE(v) + i\psi(\lambda) - 
    \ip{E}{E}_{V^{*}} - \ip{\psi}{\phi(\psi)}_{V^{*}}
    \label{eq:philocalization1}
\end{equation}
(remembering to pick up a minus sign when we move \( Q \) past the \( 
\psi \) in the second term).  Now we compute 
\[ U(\phi, v, \lambda) = \frac{1}{(2\pi)^{2\dim V}}
\int_{V^{*}\times \Pi V^{*}}d\mathrm{vol}(E)\,d\mathrm{vol}(\psi)
e^{\Phi_{L}}. \]
We now use the fact that Gaussian integration gives
\begin{align}
    \int_{V}d\mathrm{vol}\,e^{-\ip{v}{Av} + \ip{B}{v}}
    &= \int_{V}d\mathrm{vol}\,e^{-\frac{1}{4}\ip{B}{A^{-1}B}}
    e^{-\ip{v-A^{-1}v}{A(v-A^{-1}v)}} \\
    &= e^{-\frac{1}{4}\ip{B}{A^{-1}B}} \left(
    \frac{\pi}{\det A}
    \right)^{\frac{\dim V}{2}}
    \label{eq:gaussian}
\end{align}
and obtain
\begin{equation}
    U(\phi, v, \lambda) = \frac{1}{(4\pi)^{\dim V}} \int_{\Pi
    V^{*}}d\mathrm{vol}(\psi)\,e^{-\frac{1}{4}\ip{v}{v}_{V} 
    + i\psi(\lambda)
    -\ip{\psi}{\phi(\psi)}_{V^{*}}}
    \label{eq:philocalization}
\end{equation}
\gc{Am I off by a minus sign on that third term in the exponential?}

To obtain an element of the Cartan algebra, we use the fact that \( 
\Phi_{L} \) is a linear function of \( \lambda \) 
and so can be identified with a 1-form on \( V \).
If we now choose a connection \( a \) on \( A \) then we can construct
the map \( w:S^{*}(\g^{*})\otimes\formsij{*}{V}\to\formsij{*}{P\times 
V} \) by
sending \( \phi\to F_{a} \).  This is the \emph{Weil homomorphism}. 
It is an equivariant map because \( F_{a} \) transforms in the adjoint 
representation, and so descends to a map on \( G \)-invariant forms
\[ \bar{w}:\left(S^{*}(\g^{*})\otimes \formsij{*}{V}\right)^{G} \to 
\formsij{*}{E}. \]
The form \( w(U) \) is almost a representative of the Thom class 
\thom.  In fact, \( w(U) \) fails to be fully horizontal, which is
required for it to be the lift of a form from \( E \).  Use the
connection to decompose \( T_{(x,v)}(A\times V) \) into \(
T_{x,\mathrm{vert}}\oplus T_{x,\mathrm{horiz}}\oplus V \) and to form 
the projection \( p \) onto \( T_{x,\mathrm{horiz}}\oplus V. \)  
Then we denote
\[ w(U)_{\mathrm{horiz}}(X_{1},\ldots,X_{n}) = 
   w(U)(p(X_{1}),\ldots,p(X_{n})).\]
This horizontal element does in fact descend to \formsij{*}{E}.
The fact that \( U \) was already \( G \)-invariant but not 
horizontal perhaps indicates that the construction is better off 
living in the \emph{Weil model} of equivariant cohomology, but we 
follow standard practice and take a horizontal projection.  Note that 
we never had to use \( \bar{w}, \) since we projected horizontally 
after applying \( w \).  

\begin{theorem}[Mathai-Quillen \cite{mathaiquillen}] \( 
    w(U)_{\mathrm{horiz}} \) is a representative for the Thom 
class 
    \thom.
\end{theorem}

Heuristically, we see that the Berezinian integration over \( \psi \)
will give us the Pfaffian of \( F_{a} \), just as in the Gauss-Bonnet
formula.  The Gaussian in \( v \) and the constants ensure the
integral over a fiber is 1.  We obtain a top-dimensional form in the
\( V \) direction because of the \( i\psi(\lambda) \) term and the 
correspondence between functions of \( \lambda \) and forms.

Now we get back to the point about \( v \) and \( \lambda \).  The
Thom class can be pulled back by a section \( s:B\to E \) to produce
the Euler class of \( E \).  This is an \( n \)-form on \( B \)
(recall that \( \dim V = n \)), unless \( n>\dim B \) in which case 
the
Euler class is defined to be zero.  We can pull the Mathai-Quillen
form back to \( B \) by \( s \) to obtain
\begin{equation}
    s^{*}U(\phi, v, \lambda) = \frac{1}{(4\pi)^{\dim V}} 
\int_{\Pi
    V^{*}}d\mathrm{vol}(\psi)\,e^{-\frac{1}{4}\ip{s}{s}_{V} 
    + i\psi(ds)
    - \ip{\psi}{\phi(\psi)}_{V^{*}}}
\end{equation}
where we simply replaced \( v \) by \( s \) and \( \lambda \) by \(
ds \) to effect the pullback.  This form represents the Euler class 
of \( E \).  

If we replace the section \( s \) by \( ts \) for a real parameter \( t 
\) and rescale \( \psi \) by \( \psi\to \frac{1}{t}\psi \) then this 
expression becomes
\begin{equation}
    s^{*}U(\phi, v, \lambda) = \frac{1}{(4\pi)^{\dim V}} 
\int_{\Pi
    V^{*}}d\mathrm{vol}(\psi)\,e^{-\frac{1}{4t^{2}}\ip{s}{s}_{V} 
    + i\psi(ds)
    - t^{2}\ip{\psi}{\phi(\psi)}_{V^{*}}}.
    \label{eq:mqpulledback}
\end{equation}
This version allows us to consider the two limits \( t\to 0 \) and \( 
t\to\infty \) that link the Gauss-Bonnet formula with a formula that 
involves local data at the zero set of \( s \).

We will now see that our action on \sa\ has an Euler class part that is 
the pullback by \( F_{A}^{+}:\bb\to\aa\times_{\gg}\Omega^{2}_{+}(X; 
\adp). \) 
Under the identification of \( \Omega_2^+ \) with \( V^* \) and 
\lieg\ with \g, we identify \( E_2 \) with \( E \), \( \psi_2 \) with 
\( \psi \) and \( \phi \) with \( i\phi \).  So \( L(\aa) \) is an 
analogous space to the one we were just considering.  Now let us 
compare \eqref{mqpulledback} with the action \eqref{twistedaction}.  
If we choose the section \( s \) to be the map
\[ \frac{1}{2}F_A^+:\aa\to\aa\times\Omega_2^+(X; \adp) \]
then \( ds \) will take \( \psi_1 \) to \( d_A^+\psi_1 \).  Also, the 
action of \lieg\ on \(\Omega_2^+ \) is \( \psi_2\mapsto 
[\phi,\psi_2]\) and so the analogy with \eqref{mqpulledback} gives
\begin{equation}
   \label{eq:lpartofpathintegral}
   (tF_A^+)^*U = \int_{\Pi(\Omega_2^+(X; \adp))} d\psi_2\,
   e^{-\frac{t^{2}}{4}|F_A^+|^{2}+i\ip{\psi_2}{d_A^{+}\psi_1} -    
\frac{1}{t^{2}}\ip{\psi_2}{[\phi,\psi_2]} }
\end{equation}
which is part of \eqref{twistedaction} with \( t \) replaced by \( 1/g 
\).  So not only does part of the twisted action represent the 
Poincar\'e dual of the ASD moduli space, but the physical coupling 
constant plays an analogous role to the scale of the section \( 
F_{A}^{+} \)!  We will dwell on this after discussing the projection 
form.
\subsubsection{The projection form}

The construction of the projection form follows in the same vein as 
the Mathai-Quillen construction above.  We will introduce a space 
together with a vector field.  We will differentiate a function to 
obtain another function that we then exponentiate as before, and 
integrate over some of the variables.  This form will interact with 
the Mathai-Quillen form, as they will share certain variables.  In 
fact, the projection form will enforce both of the modifications we 
made to \( U \) above: it will kill off all but the horizontal 
part of \( U \) and it will produce a \( \delta \)-function that 
is supported where \( \phi \) is equal to the curvature of the chosen 
connection (the construction of the projection form involves a choice 
of connection).

Let \( A\to B \) be a principal bundle with group \( G \).  Let \(
\ip{}{}_{\g} \) be an bi-invariant inner product on \g.  Suppose we
are given a \( G \)-equivariant metric \( g \) on \( A \).  Then \( g
\) induces a natural connection on \( B \) by using the metric to take
the horizontal distribution to be the orthogonal complement of the
vertical one.  The action of the group \( G \) on \( A \) induces a
map from \g\ to the vertical tangent spaces of \( A \).  We call this
map \( C, \) so we have \( C:\g\to T_{a}A. \) Using the metric, we can
define the adjoint \( C^{*} \) of \( C \), \( C^{*}:T_{a}A\to\g \). 
In other words, \( C^{*} \) is a Lie algebra-valued 1-form on \( A \).

We will examine the space
\[ P(A) = \g\times A\times\Pi \Omega^{1}(X; \adp)\times\g\times\Pi\g \]
with the vector field
\begin{equation}
    Q'\left(
    \begin{array}{c}
	\phi \\
	a \\
	\psi' \\
	\bar{\phi} \\
	\bar{\psi} 
    \end{array}
    \right) = \left(
    \begin{array}{c}
	0 \\ \psi' \\ -L(\phi)a \\ \bar{\psi} \\ -L(\phi)\bar{\phi}
    \end{array}
    \right)_{(\phi,a,\psi',\bar{\phi},\bar{\psi})}.
    \label{eq:qprojection}
\end{equation}
We begin with the element 
\begin{equation}
    \Psi_{P}(\phi,a,\psi',\bar{\phi},\bar{\psi}) = 
    i\ip{\bar{\phi}}{C^{*}}_{\g}.
    \label{eq:psiprojection}
\end{equation}
We won't be able to compute \( Q'\Psi \) with just 
\eqref{qprojection}, 
though.  This is because \( C^{*} \) is a genuine 1-form, not an odd 
object.  However, by \propref{qinducescartan} we can work with
the Cartan differential which operates by \( (d-i(\phi))C^{*} = 
dC^{*} - C^{*}(C\phi). \)  And so we have
\begin{equation}
    \Phi_{P} = Q'\Psi_{P} = i\ip{\bar{\psi}}{C^{*}}_{\g} 
    + i\ip{\bar{\phi}}{dC^{*}}_{\g}
    - i\ip{\bar{\phi}}{C^{*}(C\phi)}_{\g}.
    \label{eq:phiprojection1}
\end{equation}
The projection form is then 
\begin{equation}
    U'(\phi, a, \psi') = \frac{1}{(2\pi i)^{\dim 
G}}\int_{\g\times\Pi\g} 
    d\mathrm{vol}(\bar{\phi})\,d\mathrm{vol}(\bar{\psi})\,e^{
    i\ip{\bar{\psi}}{C^{*}}_{\g} 
    + i\ip{\bar{\phi}}{dC^{*}}_{\g}
    - i\ip{\bar{\phi}}{C^{*}(C\phi)}_{\g}}
    \label{eq:phiprojection}
\end{equation}
which lies in \( S^{*}(\g^{*})\otimes\formsij{*}{A}. \)

\begin{prop}
    Let \( \omega \) be an element of \( S^{*}(\g^{*})\otimes 
    \formsij{*}{V} \) for some vector space \( V \) with \( G 
    \)-action. Then 
    \( \int_{\g} d\phi\, \omega\wedge U' = w(\omega)_{\mathrm{horiz}} \) 
\end{prop}
\proof For a more detailed treatment of \( U' \), see Section~14.3.3
of \cite{cordes}.  The Berezinian integral over \( \bar{\psi} \)
picks off the piece of maximal degree in \( \bar{\psi} \).  If \( \dim
G = m \) then this yields an \( m \)-form built from the \( m \)-fold
wedge product of \( C^{*}.  \) Since \( C^{*} \) is a vertical 1-form
(it vanishes on horizontal vectors in \( TA \)) this wedge product is
an \( m \)-form along strictly vertical directions.  In fact, it is an
element of \( \wedge^{\mathrm{top}}(T^{*}_{\mathrm{vert}}A).  \) Any
form on \( A \) with components along vertical directions is zero when
wedged with such a fully vertical form, and so multiplying \( \omega
\) by \( U' \) picks off the horizontal part of \( U' \).

Next, we note that the integral over \( \bar{\phi} \) in \( U' \) 
gives 
the \( \delta \)-function
\[ \delta(dC^{*}-C^{*}(C\phi)) \]
which is zero unless
\[ \phi = (C^{*}C)^{-1}dC^{*}. \]
\begin{lemma}
    \( (C^{*}C)^{-1}dC^{*} \) is the horizontal part of the curvature
    two-form \( F \) on \( B \) induced from the connection induced
    from the metric \( g \) on \( A \).
\end{lemma}
\proof Let \( X, Y \) be two horizontal vector fields on \( A \).  
The curvature two-form computes the opposite of the vertical part 
of the Lie bracket of \( 
X \) with \( Y \).  \( C^{*} \) vanishes on horizontal vectors and so
\[ dC^{*}(X, Y) = XC^{*}(Y) - YC^{*}(X) - C^{*}([X,Y]) = 
-C^{*}([X,Y]). \]
Composing this with \( (C^{*}C)^{-1} \) gives the curvature's vertical 
projection, which actually lives in \g, but can be brought back to a 
vertical vector field on \( A \) by composing with one more \( C \)  
to give
\[ C(C^{*}C)^{-1}C^{*} \]
which clarifies that this is a projection.  This completes the proof
of the Lemma which implies the Proposition.

Now let us examine the analogous objects over \aa\ and again compare 
with \eqref{twistedaction}. We need to do a couple of computations 
first to get the right expressions.  Here we follow 
\cite{atiyahjeffrey}.  First, the operator \( C^* \) is a 
\lieg-valued 1-form on \aa, and a standard formula from gauge theory 
gives for a connection \( A \)
\[ C^*(A)(\eta) = d_A^*(\eta) \]
where \( \eta\in\formsij{1}{X;\adp}. \) Similarly, the map \( C \), 
which is a map from \lieg\ into \( T\aa \) is given by
\[ \phi \mapsto d_A\phi \]
for \( \phi\in\formsij{0}{X;\adp}\cong\lieg \).

What about the map \( dC^* \)?   This will be a 2-form on \aa, and we 
can argue as follows.  In finite dimensions, if a 1-form is given by 
\( \sum f_i(x)dx^i \) then 
\[ d\left(\sum f_i(x)dx^i\right) = \sum \frac{\partial f_i}{\partial 
x^j}dx^j\wedge dx^i, \]
so what we are looking to do is differentiate \( d_A^* \) in the \( A 
\) direction. Let \( \bar{\phi}\in\lieg. \) Consider the expression 
\(\ip{\bar{\phi}}{C^*} \).  On an element \(Y_1\in T\aa\) this 
function gives
\[ \ip{\bar{\phi}}{d_A^* Y_1} = \ip{d_A\phi}{Y_1}. \]
Differentiating this in the direction of the tangent vector \(Y_2\) 
gives
\[ \ip{[Y_2,\bar{\phi}]}{Y_1}. \]
The invariance of the metric under the adjoint action of \lieg\ 
implies
\[ \ip{[Y_2,\bar{\phi}]}{Y_1} = \ip{\bar{\phi}}{[Y_1,*Y_2]}. \]
This is the 2-form \ip{\bar{\phi}}{dc^*} evaluated on \(Y_1\) and 
\(Y_2\), and now we wish to express this as a quadratic function on 
\(\Pi T\aa\).  If \(\psi_1\in\Pi T\aa\) then the corresponding 
function is just
\begin{equation}
   \label{eq:2formasfunction}
   \ip{\bar{\phi}}{[\psi_1,*\psi_1]}.
\end{equation}
And so the analog of the projection form is
\begin{equation}
   \label{eq:ppartofpathintegral}
   U' = \int_{\lieg\times\Pi\lieg}D\bar{\phi}\,D\psi_0\,
    e^{i\ip{\psi_0}{d_A^*\psi_1} + i\ip{\bar{\phi}}{[\psi_1,*\psi_1]} 
-     
    i\ip{\bar{\phi}}{d_A^*d_A\phi}}
\end{equation}
which forms another part of the action \eqref{twistedaction}.

We have not discussed two of the terms in \eqref{twistedaction}. 
Those are \( -i|E_{2}|^{2} \) and \( i\ip{\phi}{[\psi_{0},\psi_{0}]}. 
\) These two terms are not used in the analogy with geometry that we
have just constructed, but nor do they pose a problem.  In fact, if one
enforces the classical equations of motion for the auxiliary field \(
E_{2} \) one obtains \[ -i|E_{2}|^{2}\to -i|[\phi,\bar{\phi}]|^{2}, \]
which we will not prove.  The quantity \[ -i|[\phi,\bar{\phi}]|^{2} +
i\ip{\phi}{[\psi_{0},\psi_{0}]}\] is in the image of a Cartan
differential, just as the rest of the action was shown to be above. 
However, in this case passing to equivariant cohomology actually kills
off these two terms, and so they are not important to our story.  The
\( [\phi,\bar{\phi}] \) term is of crucial importance in studying the
classical and quantum vacua of the physical theory on flat 
space.

\subsection{Path integrals}
Consider a path integral of the form 
\[\int_{\mathrm{fields}}e^{-S_{q}+\sum_{i}g_{i}S_{I,i}} \] where the
various \( S_{I, i} \) are \emph{interaction terms}, which just means
they are each a cubic or higher order function on field space.  \(
S_{q} \) is a quadratic function on field space.  Such an integral can
be written as a formal power series in the \( g_{i} \), the coupling
coefficients.  This series does not converge, and each term in the
series diverges unless we renormalize.  So what we're dealing with
here is no better off mathematically than the Thom class idea is. 
However, the constant term (independent of all \( g_{i} \)) is
computed by evaluating only the quadratic part \( S_{q} \) of the
action, which is a Gaussian integral that can be rigorously defined
using zeta function regularization of determinants and Pfaffians of 
infinite-dimensional operators.  So,
in the limit as the coupling parameters vanish, the quadratic part is
the whole of the path integral.  This is a \emph{free theory}, which 
means it models a theory of particles that do not interact with each 
other.  

Our theory has one coupling parameter, \( g \) and so to compute the
free path integral we'd be taking the limit \( g\to 0 \). 
Furthermore, if an action has a moduli space of minima, one computes
terms in the perturbation expansion by integrating over this moduli
space and projecting the path integral onto the normal bundle of this
space.  That would be how we'd compute the Donaldson invariants as
path integrals, too.  The quadratic part of the action acting on the
normal bundle to \( \mm\in\bb \) has quadratic part consisting of the
Laplacian \( d_{A}^{*}d_{A} \) on even objects and \( d_{A} \) or \(
d_{A}^{*} \) on fermions.  When carefully computed, the resulting
determinants and Pfaffians of these operators will cancel up to sign,
the details of the sign depending on considerations involving the
orientation of \mm.

Something very special is happening here, though.  The reciprocal of
the coupling parameter plays the same role as the scale of the section
\( F_{A}^{+} \), as we saw when we compared \eqref{twistedaction} and
\eqref{lpartofpathintegral}.  And so low coupling corresponds to
taking the scale to infinity.  The algorithm we use at low coupling to
compute the path integral aligns exactly with the steepest descent
computation one uses to show that the Mathai-Quillen form can be 
expressed in terms of local data on the zero set of the section.  

Atiyah and Jeffrey made a related statement in \cite{atiyahjeffrey}. 
They pointed out that the Mathai-Quillen construction could allow the
definition of a \emph{regularized} Euler class.  Even if the base
space and vector bundle are infinite-dimensional, if we choose a
section that has a finite-dimensional zero set \( M \), then we can
define the regularized Euler class in terms of \( M \).  This is
similar to saying that one can define a path integral at zero
coupling.  Perhaps results in either the path integral direction or 
the Thom class direction can inform the other.

\subsection{The quantum observables \mytexorpdfstring{\oi{i}}{O(i)}}

If we want to do intersection theory on the zero set of \( F_{A}^{+} 
\) then we are all ready, because we have an Euler class to wedge 
forms 
against, which is equivalent to integrating the forms over the 
Poincar\'e dual of the Euler class, which is of course exactly the 
zero set in question.  To do this within the field theory framework, 
we need superspace representatives of the forms used in Donaldson 
theory.  These involve slant products with a Pontrjagin class, and so 
we will find a field theory representative for this construction.

The following is a standard construction from Donaldson theory.  See
\cite{donaldsonkronheimer} for more details.  Let \( P\to X \) be a
principal \sut\ bundle over an even riemannian four-manifold \( X \). 
We denote by \aa\ the space of connections and by \gg\ the group of
gauge transformations.  Let \( \aa^{*} \) and
\( \bb^{*} \) be the respective complements of the set of reducible
connections.  Let \( \bb^*=\aa^*/\gg.  \) The bundle
\[ \PP=\aa^{*}\times_{\gg}P\to \bb^{*}\times X \] 
is a principal \sot\ bundle, and so has a first Pontrjagin class \(
p_{1}(\PP) \).  We define a connection on this bundle by using a 
metric as follows.  Give \( \aa\times\{p\} \) 
the usual metric on \aa\ and give \( \{A\}\times P \) a metric by 
using the metric on \( 
X \) for horizontal vectors, and the connection \( A \) together with 
a metric on \lieg\ for vertical vectors.
There is an associated connection given by taking the orthogonal
complements of the vertical subspaces of \( T(\aa^{*}\times P).  \)
The curvature of this connection, \ff, is given by the following
formulas.  Let \( \tau_{1} \) and \( \tau_{2} \) be horizontal
tangents to \aa\ at \( A \), and let \( X_{1} \) and \( X_{2} \) be
horizontal tangents to \( P \) at \( p \).  Then one computes
\begin{align}
    \ff_{2,0}(A, p)(\tau_{1}, \tau_{2}) &= 
    -\frac{1}{d_{A}^{*}d_{A}}d_{A}^{*}([\tau_{1}, \tau_{2}]) \\
    \ff_{1,1}(A, p)(\tau_{1}, X_{1}) &= \tau_{1}(X_{1}) \\
    \ff_{0,2}(A, p)(X_{1}, X_{2}) &= F_{A}(X_{1}, X_{2}).
\end{align}
The subscripted indices denote the bigrading in \( 
H^{*}(\bb^{*}\times X). \)  The bracket \([\tau_1,\tau_2]\) is a 
bracket of two vector fields on \aa, not the bracket as sections of 
\adp.

To see how to create a superspace representation \( \FF \) of \ff, we
just need to come up with an equivariant representative of each of
these three 2-forms.  The generator \( \phi \) of \( S^{*}(\lieg^{*})
\) maps to the curvature \( \ff_{2,0} \) under the Weil map, and so
properly interpreted, \( \phi \) is \( \FF_{2,0} \).  What the
generator \( \phi \) means written alone is \emph{the identity
function} on \lieg.  This is the element 
\[ \phi\in\lieg^{*}\times\lieg\cong \lieg^{*}\times \formsij{0}{X;
\adp}.  \]
Similarly the expression ``\( \psi_{1} \),'' when written in 
isolation, is an identity function, this time on \( \Pi T\aa \).  
That makes \( \psi_{1} \) a vector-valued function on \( \Pi T\aa. \) 
Under the correspondence with forms, this becomes a vector-valued 
1-form corresponding to the identity function on tangent vectors, 
otherwise known as de Rham \( d \).  So, \( \psi_{1}\leftrightarrow d 
\), which can be evaluated on a pair \( (\tau_{1}, X_{1}) \) as above 
to give
\[ d:(\tau_{1}, X_{1})\to (\tau_{1}, X_{1}) \]
which is the identity, yielding a tangent vector to the space of 
connections and a tangent vector to the manifold.  And so the final 
evaluation of \( \psi_{1} \) is
\[ \tau_{1}(X_{1}) \]
giving the identification between \( \psi_{1} \) and \( \ff_{1,1} 
\).  \( F_{A} \) 
is already a field in our theory and so we obtain
\begin{align}
    \FF_{2,0} &= \phi\\
    \FF_{1,1} &= \psi_{1} \\
    \FF_{0,2} &= F_{A}.
\end{align}

It is worthwhile to note that if we compute the action of the vector
field induced by \( Q_{1} \) on \( \phi \) we obtain
\begin{equation}
\tighten{
\begin{array}{lcl}
    \qoc\phi &=& \psi_{1} \\
    \qoc\psi_{1} &=& F_{A} + \qzc(F(d,D_{1}))
\end{array}}
\end{equation}
so that if we compute modulo \qzc\ (i.e. we work on the level of 
equivariant cohomology and not equivariant forms) then \qoc\ 
permutes bigraded pieces of \FF, keeping total grading invariant.

The Pontrjagin class \( p_{1}(\PP) \) is given by 
\[ p_{1}(\PP) = \frac{1}{2}\mathrm{Tr}(\ff\wedge\ff), \]
and so we will examine the equivariant representative
\[ \frac{1}{2}\mathrm{Tr}(\FF\wedge\FF). \]
To compute the slant product with an \( i \)-dimensional homology 
class, 
we just integrate a bigraded piece over a smooth representative. So 
Donaldson's \(\mu\)-map is
\[ \mu([\Sigma_{i}]) = p_{1}(\PP)/[\Sigma_{i}] = 
   \int_{\Sigma_{i}} \left(\frac{1}{2}\mathrm{Tr}
   (\FF\wedge\FF)\right)_{4-i,i}. \]
For the record we list the bigraded pieces
\begin{align}
    p_{1}(\PP)_{4,0} &=\frac{1}{2} \mathrm{Tr}(\phi^{2}) \\
    p_{1}(\PP)_{3,1} &= \mathrm{Tr}(\phi\psi_{1}) \\
    p_{1}(\PP)_{2,2} &= \mathrm{Tr}\left(\frac{1}{2}\psi_{1}\wedge 
    \psi_{1} + \phi F_{A}\right) \\
    p_{1}(\PP)_{1,3} &= \mathrm{Tr}(\psi_{1}\wedge F_{A})\\
    p_{1}(\PP)_{0,4} &= \frac{1}{2}\mathrm{Tr}(F_{A}\wedge F_{A})
\end{align}
\subsection{The path integral formulation of the polynomial invariants}

We already know that the projection form part of the path integral 
can be evaluated formally and gives the Weil homomorphism from 
equivariant cohomology to usual cohomology, by mapping \( \phi \) to 
the curvature of a connection.  And so we may use supersymmetric 
equivariant 
representatives for \( p_{1} \) in the construction of a path 
integral, knowing that they will be mapped to the real thing.  Thus 
we can construct a completely physical analogue of Donaldson theory, 
by integrating local operators over a space of fields, against the 
exponential of an action that is also built from local fields.
Let \( \oi{i} = p_{1}(\PP)_{4-i,i} \) be the equivariant 
representatives and let \( D \) be the 
Donaldson polynomial on second homology. We have hopefully motivated 
the following claim.
\begin{claim} The gaussian approximation to the path integral 
    \[
        \int_{\sa}\left(\prod_{k=1}^{N}
	\int_{\Sigma_{i_{k}}}\oi{i_{k}}\right)
	\exp(-S) 
    \]
    agrees with the Donaldson polynomial
    \[
    D(\Sigma_{i_{1}}, \ldots, \Sigma_{i_{N}}) = 
    \int_{\mm}\mu([\Sigma_{i_{1}}])\wedge\cdots\wedge
    \mu([\Sigma_{i_{N}}]).
    \]
\end{claim}

\section{Outlook: Witten's conjecture}
This work is the beginning of a program.  The goal is to prove
Witten's conjecture \cite{witten94}.  The physical insight that
allowed Witten to make this remarkable conjecture comes from his
celebrated paper with Seiberg \cite{seibergwitten1}.  My hope is that
one day a mathematical version of that paper may be created, and this
work marks the humble beginnings of that project.

It may be possible to prove Witten's conjecture using nonabelian
monopoles (\cite{pidstrigatchtyurin}, \cite{feehanleness},
\cite{feehanleness2}).  However, a proof that parallels the physical
proof would have the advantage that it may reveal a broader picture of
math/physics interaction, and help us to understand why so much recent
mathematics has grown out of physics.  In particular, it may shed
light on either the mathematical relevance or the mathematical
underpinnings of the renormalization group.

However, for that we need more mathematics.  We have tried to cast
Donaldson theory in physical language, along the lines that Witten
hinted at in 1988 \cite{wittentqft}.  That was the starting point for
Seiberg and Witten's breakthrough work of 1994 \cite{seibergwitten1},
and so it needs to be the starting point for the mathematical proof as
well.  Hopefully after reading this paper this picture of Donaldson
theory seems somewhat natural from a mathematical standpoint.  From a
physical standpoint, it is simply a supersymmetric gauge theory, and
the Minkowski space version (the ``physical'' theory, as it's known,
as opposed to the twisted theory, which is called ``topological'' or
``cohomological'') can be treated with all the machinery of modern
physics.  As of 1994, there wasn't enough physics to understand this
theory any better than we have already done in this paper.  However,
Seiberg and Witten created new physics that solved the theory.  To
understand what that means, we need to discuss energy scales.

\subsection{Energy scale and the renormalization group}
We have discussed the coupling parameter that appears in both the 
physical and twisted gauge theory \actions.  We talked about how the 
constant term in the coupling expansion (the perturbation series) is 
the appropriate Donaldson invariant.  We did not address what 
controls the value of the coupling constant or whether it is simply 
arbitrary.  In fact
\begin{claim}
    Let \( \gamma_{t} = e^{-t}\gamma \) be a one-parameter family of
    metrics on \( E^{4} \).  Then as \( t \) blows up, the coupling
    parameter \( g \) responds by shrinking to zero.  In other words,
    \( g \) is a function of \( t \) and we have \[ \lim_{t\to
    \infty}g(t) = 0.  \]
\end{claim}
This follows from the \emph{asymptotic freedom} of nonabelian gauge
theories.  The significance of the metric scaling to zero is in the
fact that this is the regime of \emph{high energy}.  Using the speed
of light and Planck's constant, meters and energy units can be
converted back and forth, just as meters and seconds can be converted
using just the speed of light.  Small distances correspond to large
energy units, and so shrinking the metric to zero examines the theory
at high energy.  In this paper we have therefore examined correlation
functions at the high energy limit, since as the energy approaches
infinity, the coupling goes all the way to zero, leaving only the
Donaldson invariants.  On the other hand, the perturbation expansion,
which is borderline meaningless anyway, only stands a chance of
converging if the coupling is small, and so physicists believe that
asymptotically free theories at high energy are quite under their
control to compute with.

Physicists, however, would have us make the following analogy.  Take
some fundamental model of the basic forces of nature, like string
theory or the standard model.  These are hugely complicated
asymptotically free theories that describe the behavior of our
universe at the highest of energies and the smallest of distance
scales.  They are ``fundamental'' in that sense --- they are the
underlying physics of everything.  However, more often than not we are
interested in more mundane matters like fluid flows or planetary
mechanics.  In these cases, we are working on rather larger distance
scales, and quite tiny energy regimes.  The laws of physics are
surprisingly simpler in this sort of context, becoming things like
Kepler's laws or the Navier-Stokes equations.  These are not gauge
theories or string theories with infinitely many degrees of
freedom.  They are finite-dimensional PDEs or algebraic equations
that, while perhaps difficult to solve completely, are provincial and
comparatively easy to understand.  The conceptual force at work here
is the \emph{renormalization group}, first understood in this way by
Wilson \cite{wilson}, \cite{polchinski}.  The renormalization group is
simply the group of energy scaling.  It is the \( t \) parameter in
the above theorem.  While the underlying group is simple, its action
on a theory is not.  Somehow, as it flows from high energy to low, the
complexities of the fundamental theory are suppressed, and only a few
parameters and degrees of freedom survive at low energy.  We are not
going to try to understand the renormalization group and its workings,
but instead we are just trying to paint the picture that is in the
back of all physicists' minds.  There is no overarching description 
of the renormalization group, no actual flow that can be applied to a 
theory to get answers at different scales.  It is something that is 
ill-understood at best, though perhaps the route to addressing it and 
constructing a workable theory is via the sort of topological 
application we are discussing here!

Witten makes the analogy \cite{wittenias} between a fundamental,
asymptotically free theory, and a differential equation.  The former 
contains
infinitesimally small distance information about a theory, as we said
before, and so the analogy with information about a function's
derivatives is very close.  The ability to describe a theory at any
energy scale is, then, analogous to solving the differential
equation.  Nonabelian gauge theories are therefore candidate
``problems,'' and the challenge is to find their ``solutions.'' 
Donaldson theory is asymptotically free because it is an \sut\ gauge
theory.  Its correlation functions are the Donaldson polynomials, and
the challenge in terms of renormalization is to compute these
correlation functions at all energies, and \emph{solve Donaldson
theory}.  Mathematics has no basis for even stating this challenging
problem.  Donaldson theory is not a question at all in its
mathematical presentation.  It is simply a construction that yields
interesting information.  But physics goes further, and places it at
one end of the real numbers, near infinity, because that is
where it fits in terms of energy scale.  Of what picture is this the
limit?  No one has addressed this question.  What has been
addressed, though, is what lies at the other end of the line, at zero
energy.  The answer: Seiberg-Witten invariants.

Abelian gauge theories are the opposite of asymptotically free: the
coupling parameter vanishes for \emph{low} energy (the \( t\to 
-\infty \)
limit in the above theorem).  These are called \emph{infrared free}
theories.    Infrared free theories are, then, 
the candidates for the solutions to asymptotically free ones.  They 
are well-behaved as perturbation series only near zero energy.  
Seiberg-Witten theory is a theory of a \uo\ gauge field coupled to a 
spinor, and so is an abelian gauge theory.  As such it is infrared 
free, and is one of a host of candidates for solutions to Donaldson 
theory.  In fact, Seiberg and Witten show that it is 
exactly the right solution.

\subsection{Overview of the physics proof}
To describe the physics proof of the conjecture would be too vast an 
undertaking for this paper, and so we merely sketch it, pointing out 
some of the deep issues that will need confronting.  First of all, 
the result does not involve the twisted theory we 
have considered here, but rather the \( N=2 \) Minkowski space 
version.  However, Witten's conjecture clearly involves assuming 
that the flow to low energy commutes with the twisting operation, as 
one simply twists the low energy physical theory to obtain the 
Seiberg-Witten invariants.  The good news then is that we have 
learned all the mathematical tools needed to twist the low energy 
physical theory in the preceding sections of this paper.  The bad news 
is that it may not be possible to cast a workable parallel of the flow 
to low energy in twisted terms.

The central issue to sort out, however, is the concept of the
\emph{quantum vacuum}.  In classical physics, the minima of the action
form the \emph{classical vacuum manifold}.  The action principal
states that a physical system will assume one of the configurations
from this space.  In quantum physics, a system can assume any of the
states in the whole field space, but the path integral weighs minima
of the action with much higher probability.  To turn the crank of
quantum field theory, one must select a classical vacuum and write the
action in coordinates that perturb from this state.  The
\emph{quantum} vacuum for this theory is then a state in the theory's
Hilbert space that has no particles, or is invariant under the entire
Poincar\'e group.  This indirect definition makes it hard to quantify,
especially because it is difficult or impossible to actually construct
this Hilbert space!  Certain details of the theory influence whether
the quantum vacuum is unique or not.  In the case of our \sut\ theory,
the classical vacuum manifold is the complex plane modulo the action
of \( z\mapsto z^2 \), with the resulting cone singularity at the
origin.  It is believed that the quantum vacuum manifold is this same
space, but with a different metric and other properties.  This is a
guess.  There is currently no way to ascertain the validity of this,
even physically.  However, if one assumes a whole lot about physical
theories and what sort of perturbations of classical vacua are
permitted to arise from quantization, then it is the simplest guess. 
There are singularities on the quantum vacuum manifold as well, but
there are two of them, and neither of them is at \( z=0.  \) The whole
of the matter revolves around proving what this manifold is, what
metric it has, and what the monodromies of the coupling parameter
around the singularities are.  Other details are important too,
however.  Seiberg and Witten claim that there are ``BPS states'' in
each theory of the quantum manifold, and that the mass of these states
is different in each theory.  The two singularities are the two points
where this mass vanishes.  That is what is making the metric blow up
at those points, they claim.  A BPS state is by definition a vector in
the Hilbert space that is annihilated by \emph{half} of the eight
supersymmetry operators (which act on the Hilbert space), so this
object, which is some sort of soliton, should be quite tractable
mathematically.  Perhaps this is the way to access a mathematical
theory about quantum vacua and the conjecture.


\bibliographystyle{unsrt}
\bibliography{SUSYDonaldsonLangmead}
\end{document}